\documentclass[journal]{IEEEtran}

\usepackage{xcolor,soul,framed} 
\usepackage{amsfonts} 
\usepackage{multicol}
\usepackage{multirow}
\usepackage{booktabs}
\usepackage{hyperref} 
\colorlet{shadecolor}{yellow}
\usepackage[pdftex]{graphicx}
\graphicspath{{../pdf/}{../jpeg/}}
\DeclareGraphicsExtensions{.pdf,.jpeg,.png}
\usepackage[numbers,sort&compress]{natbib}
\usepackage[cmex10]{amsmath}
\usepackage{array}
\usepackage{mdwmath}
\usepackage{subfigure}
\usepackage{lscape}
\usepackage{mdwtab}
\usepackage{eqparbox}
\usepackage{url}
\usepackage{float}
\usepackage{bm}
\usepackage{lineno}
\usepackage{makecell}
\usepackage{url}
\usepackage{tikz}
\usetikzlibrary{positioning}
\usepackage{calc}
\usepackage{diagbox}
\usepackage{url}
\usepackage{threeparttable}
\usepackage{verbatim}
\usepackage[figuresright]{rotating}
\def\BibTeX{{\rm B\kern-.05em{\sc i\kern-.025em b}\kern-.08em
    T\kern-.1667em\lower.7ex\hbox{E}\kern-.125emX}}
\title{
Towards Next-Generation Steganalysis: LLMs Unleash the Power of Detecting Steganography
}
 \author{ Minhao Bai$\dagger$,
 Jinshuai Yang$\dagger$,
 Kaiyi Pang,
 Huili Wang,
      Yongfeng Huang
      \thanks{$\dagger$ Equal contribution.}
      \thanks{M. Bai, J. Yang, K. Pang and H. Wang are with the Department of Electronic Engineering, Tsinghua University, Beijing, 100084, China. (e-mail: bmh22@mails.tsinghua.edu.cn) }
      
      \thanks{Y. Huang is with the Zhongguancun Laboratory, Beijing, 100084, China. (email: yfhuang@tsinghua.edu.cn)}
    }

\begin{document}
\maketitle


\begin{abstract}
Linguistic steganography provides convenient implementation to hide messages, particularly with the emergence of AI generation technology.
The potential abuse of this technology raises security concerns within societies, calling for powerful linguistic steganalysis to detect carrier containing steganographic messages. Existing methods are limited to finding distribution differences between steganographic texts and normal texts from the aspect of symbolic statistics. However, the distribution differences of both kinds of texts are hard to build precisely, which heavily hurts the detection ability of the existing methods in realistic scenarios. To seek a feasible way to construct practical steganalysis in
real world, this paper propose to employ human-like text processing abilities of large language models (LLMs) to realize the 
difference from the aspect of human perception, addition to traditional statistic aspect. Specifically, we systematically investigate the performance of LLMs in this task by modeling it as a generative paradigm, instead of traditional classification paradigm. Extensive experiment results reveal that generative LLMs exhibit significant advantages in linguistic steganalysis and demonstrate performance trends distinct from traditional approaches. Results also reveal that LLMs outperform existing baselines by a wide margin, and the domain-agnostic ability of LLMs makes it possible to train a generic steganalysis model\footnote{Both codes and trained models are openly available in \url{https://github.com/ba0z1/Linguistic-Steganalysis-with-LLMs}}.

\end{abstract}

\begin{IEEEkeywords}
    Linguistic Steganalysis, Large Language Models, Generative Steganalysis.
\end{IEEEkeywords}

\section{Introduction}

\IEEEPARstart{S}{teganography} is the technology to hide messages within normal information carriers such as images \cite{imagestega1,imagestega2,imagestega3}, audios \cite{audiostega1,audiostega2}, or texts \cite{yang2018rnn,yang2020vae,zhang2021provably,de2022perfectly,ding2023discop,yang2023semantic,fang2017generating,ziegler2019neural,shen2020near,dai2019towards,yang2018automatically}. It allows individuals to covertly transmit secret messages by embedding them within seemingly innocent content. 
With the most advanced generative linguistic steganography, a paragraph of text can embed a secret message of several hundred bits, making it possible to transmit complex information. 
It is within these inconspicuous sentences that messages are covertly conveyed.

Due to the peculiar nature of this technology, the abuse of steganography raises security concerns within societies, leading to significant anxiety among the general public \cite{zielinska2014trends, mazurczyk2017information}. Unethical individuals may employ steganography to convey malicious information, such as terrorists using steganography to plan attacks \cite{terror911, BinLaden}, or hackers using it to distribute virus software \cite{hacker}.
Therefore, the detection of steganographic carriers containing secret messages, namely steganalysis,  has emerged as a widely discussed area in contemporary information security.


Understanding steganography is the foundation of steganalysis. Thanks to the widespread usage of texts and the extraordinary achievement of AI generation techiques, generative linguistic steganography has become a popular choice for research and practice. Generative linguistic steganography embeds secret messages into automatically generated texts. The advantage lies in the absence of a predefined carrier, resulting in enormous flexibility and a high embedding rate. In the text generation process, a codebook is formed using the conditional probability distribution predicted by a language model. To construct the codebook, researchers have tried various methods including heuristic rules \cite{fang2017generating}, source coding based methods \cite{yang2018automatically, dai2019towards, shen2020near, ziegler2019neural} and distribution preserving methods \cite{zhang2021provably,ding2023discop}. After the codebook is determined, the appropriate tokens are selected based on the secret messages.

Constructing a secure codebook is critical to the imperceptibility of steganographic texts, especially in the era that large language models can build good enough conditional probability distribution. Among these codebook construction methods, heuristic methods \cite{fang2017generating} expose significant statistical distribution difference between steganographic texts and normal texts, source coding based methods  \cite{yang2018automatically, dai2019towards, shen2020near, ziegler2019neural} can make the deviation small, while distribution preserving methods  \cite{zhang2021provably,ding2023discop} can make the deviation tiny enough to near zero.

To distinguish between the steganographic texts (also called stegos) and the natural texts (also called covers), these existing linguistic steganalysis methods depend on statistical features to be aware of the distribution deviation, acting as classifiers. The earlier methods depend heavily on heuristic statistical features \cite{chen2011steganalysis,xiang2014linguistic}, unaware of the deep features of texts. Recent approaches focus on assembling existing neural network modules to get different learning-based deep features of texts \cite{yang2019fast,TS-CSW,TS-RNN,GNN, li2022detection, guo2022linguistic, xue2022effective, wen2023scl, peng2023text}, while several recent works \cite{EILGF,zou2020high, yang2023link} choose to use pre-trained models like BERT \cite{BERT} as the base module. These approaches seriously rely on the distributional consistency between training texts and target texts, which is difficult to accurately achieve. 

\begin{figure}[t]
\includegraphics[width=0.5\textwidth]{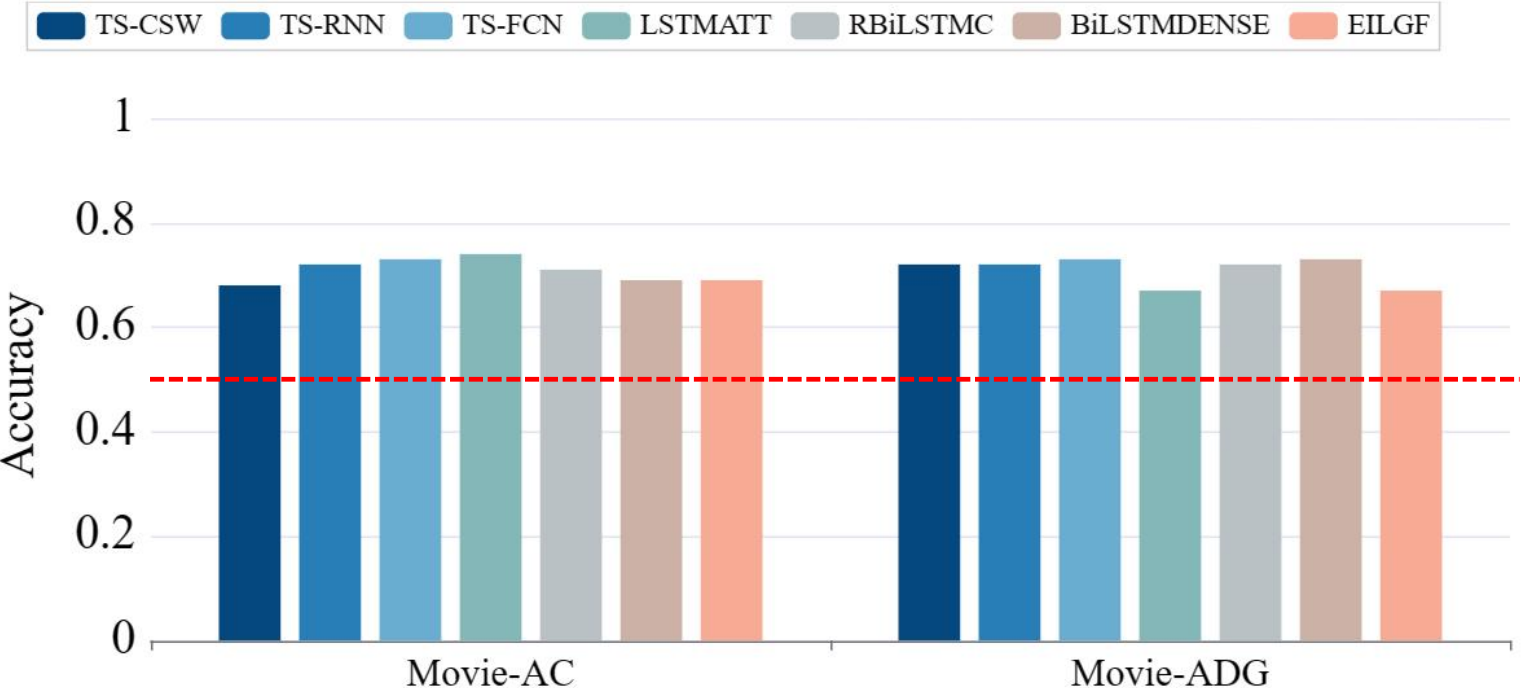}
\caption{The detection accuracy of recently  representative steganalysis methods\cite{yang2019fast,TS-CSW,TS-RNN,zou2020high, niu2019hybrid,EILGF,yang2020linguistic}. Here we test two popular codebook construction methods, namely source coding-based method AC \cite{ziegler2019neural} and distribution preserving method ADG \cite{zhang2021provably}. The red dotted line represents the 0.5 accuracy guess line.}
\label{Baselines}

\subfigure[An example of low-quality stego]
{\includegraphics[width=0.24\textwidth]{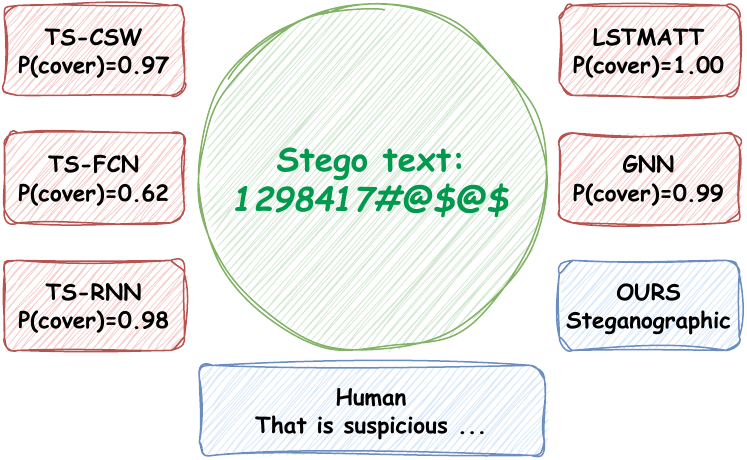}}
    \subfigure[An example of irrational stego]{\includegraphics[width=0.24\textwidth]{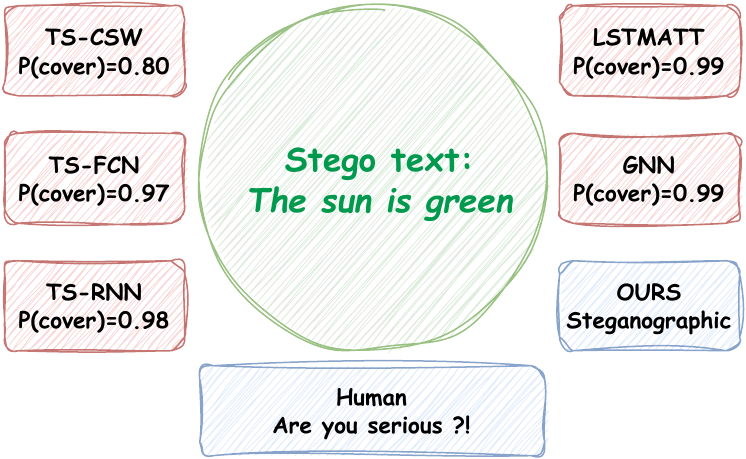}}
    \caption{Two "difficult-to-detect" cases. Current steganalysis techniques, such as TS-CSW \cite{TS-CSW}, TS-FCN \cite{yang2019fast}, TS-RNN \cite{TS-RNN}, LSTMATT \cite{zou2020high}, and GNN \cite{GNN}, produce inaccurate results with high levels of confidence for these two types of steganographic text, while our approach maintains its accuracy. In human opinion, these types of stego are unreadable or irrational, thus leading to suspicion from monitors.}
    \label{fig:cases}
\end{figure}

For heuristic codebook construction methods, it is easy for existing steganalysis methods to detect steganographic texts. However, when these recent methods meet source coding-based method and distribution preserving methods, especially the latter, they can only find little distribution difference and then show weaker detection performance. As shown in Fig. \ref{Baselines}, the detection accuracy of these existing methods is below 75\%, leaving a room for further improvement. 

There are two key reasons prevent these methods from accurately perceiving the potential distribution differences in realistic world. Firstly, building precise distribution of stegos demands massive annotated text. Secondly, pinpointing the distribution difference demands oracle distribution of normal texts in target scenario. The two excessive demands force these methods relied on distribution difference to only obtain unsatisfactory performance. 



Actually, there exists the contradiction between statistical distribution differences and text perception rationality for almost all the codebook construction methods \cite{yang2020vae}. This indicates that rationality of text deteriorates as distribution differences decrease, which suggests a clue for steganalyzers. Although evaluating rationality of text is not very difficult for humans, almost all the traditional steganalysis methods trained as statistical classifiers seem to lack this ability. As shown in Fig. \ref{fig:cases}, current steganalysis methods are easily fooled by low-quality stegos and irrational stegos, while these stegos can be distinguished by human. However, LLMs have shown human-like abilities in many aspects, including evaluating fluency and rationality of text \cite{chen2023textualityevaluationbyllm}. For example, ChatGPT has been widely used to annotate high-quality text, replacing human annotations.

Based on the above observations, a natural thought emerged: we expect that the human-like capabilities of LLMs can help steganalyzers to achieve more accurate detection, thus overcoming the performance dilemma of current linguistic steganalysis methods.
However, exploiting the capabilities of LLMs for linguistic steganalysis poses significant challenges and costs. On the one hand, activating the capability of LLMs is not easily available. Researchers have explored the potential of using ChatGPT for steganalysis tasks \cite{explore}, which shows mediocre performance due to lack of focused design and training. On the other hand, due to the vast parameters of modern LLMs, direct training requires significant time and financial investment, with the risk of losing its original capabilities. 

To seek a feasible way to construct practical steganalysis in realistic world, this paper systematically explores the LLMs for linguistic steganalysis. To hold human-like ability of LLMs as possible, we propose to generate readable detection results.
To fully activate the domain-specific steganalysis capability of LLMs, we fine-tune LLMs with various instructions, while we equip LLMs with domain-agnostic steganalysis capability by exploring appropriate mix-up of various prevalent stego texts. To achieve acceptable training costs, we adopt the lightweight training strategy of LLMs, which only requires a single GPU. Extensive experimental results show that after lightweight fine-tuning LLMs obtain the state-of-the-art linguistic steganalysis performance, in both domain-specific and domain-agnostic scenarios. 
We conduct extensive experiments on exploring the LLMs for linguistic steganalysis, and we find: 
\begin{itemize}
    \item The ability to detect stegos is just embedded in LLMs and fine-tuning can activate this ability. LLMs with instruction fine-tuning achieve much higher accuracy in detecting steganographic text than using the steganalysis model with BERT, while the trainable parameters are much less than with BERT. 
    \item Based on extensive pre-training datasets, our models exhibit exceptional detection capabilities across diverse steganographic algorithms and datasets. The proficiency stems from the LLMs' capacity to evaluate the fluency and rationality of the text. 
    \item Leveraging the cross-domain capability of LLMs, a general steganalysis model is trained on a blender of datasets. In most scenario, the performance of our general model on unseen dataset is superior to baseline methods trained on this dataset.
\end{itemize}


\section{Related Work}
\subsection{Linguistic Steganography}

Linguistic steganography can be classified into three primary categories: retrieval-based, modification-based, and generation-based linguistic steganography.

For retrieval-based text information hiding methods, such as \cite{CoverlessTextInformationHidingMethodBasedonWordRankMap} and \cite{coverless2017}, the fundamental procedure includes encoding samples from a text corpus and subsequently choosing relevant sentences for transmission, depending on the secret messages to be embedded.
Nevertheless, the need for pre-shared codebooks among communicating parties restricts its capacity and reduces its flexibility.

The fundamental concept of modification-based linguistic steganography involves making subtle changes to a given text by using secret messages as control signals. Techniques like synonym substitution \cite{bergmair2004towards} and sentence paraphrasing \cite{atallah2001natural} have been explored. 

In recent years, generative linguistic steganography methods can be classified into the following categories:
 
A. Coding methods based on rules, such as the early approach \cite{fang2017generating}, involve designing predefined rules to directly encode tokens. However, these methods solely rely on rule design and lack theoretical proof for statistical imperceptibility.

B. Coding methods that modify the distribution, such as truncation. This method first disrupts the conditional probability distribution and then source coding techniques like fixed-length coding (FLC) \cite{yang2018rnn}, Huffman coding (HC) \cite{yang2018automatically, dai2019towards}, and arithmetic coding (AC) \cite{shen2020near, ziegler2019neural} are applied to the disrupted distribution. These methods greatly undermine the statistical covertness of the steganographic text as they disturb the distribution.


C. Coding methods that preserve the distribution aim to ensure provable imperceptibility under KL divergence by achieving consistency in the distribution between steganographic and natural text. To achieve this goal, researchers have considered distribution shifting \cite{ding2023discop}, entropy coupling \cite{de2022perfectly}, grouping \cite{zhang2021provably}, and mapping \cite{shen2020near}.


\subsection{Linguistic Steganalysis}

The first-generation steganalysis methods were primarily based on heuristic statistical features. These studies \cite{chen2011steganalysis,xiang2014linguistic} focused on modification-based steganography methods, such as synonym substitution techniques that were popular at that time.
However, these methods became less effective with the continuous improvement of language models.

In recent years, second-generation steganalysis methods \cite{TS-CSW,TS-RNN,yang2019fast,zou2020high,xue2022effective,wen2023scl,guo2022linguistic,yang2023link,peng2023text} have focused on generative linguistic steganography. These methods employ deep learning models to differentiate between steganographic and natural texts using statistical features \cite{yang2019fast}. The attention mechanism is utilized in \cite{yang2019fast} to extract the relationship between word embeddings, followed by the use of a classification network with a single activation unit to output the classification results. TS-CSW \cite{TS-CSW} proposes a model based on convolutional sliding windows, which can extract correlation features using multiple window sizes. Additionally, TS-RNN \cite{TS-RNN} observes that the conditional probability distribution of each word in steganographic texts becomes distorted after secret messages are embedded. To address this, the method proposed in \cite{TS-RNN} employs recurrent neural networks to extract these distribution differences and subsequently classify texts as non-steganographic or steganographic.

Due to the limited classification ability of simple neural networks, subsequent studies adopted increasingly complex deep learning models. A model based on a graph neural network (GNN) was proposed by \cite{GNN}, where texts were transformed into graphs. In these graphs, nodes represented words, and edges captured associations between them. EILGF \cite{EILGF} focused on extracting and integrating local semantic features and global long-term dependencies to enhance the quality of text representation. The improved representation facilitated more accurate classification. \cite{peng2023text} involved fine-tuning a BERT extractor through the hierarchical supervised learning that combines signals from multiple softmax classifiers, rather than relying solely on the final one. \cite{wen2023scl} combined the cross-entropy loss and the supervised contrastive loss to guide model training to obtain better results.  

With the development of text steganography techniques, differentiating statistical features between steganographic and natural texts has become more challenging. This difficulty arises due to the existing deep learning models' struggle to capture the intricate and overlapping boundaries between the two types of text. Moreover, as LLMs can now generate text that is highly similar to natural text, the effectiveness of current deep learning methods is gradually declining.

The potential solution lies in constructing the third-generation steganalysis methods based on LLMs, leveraging their powerful capabilities to develop a specialized model in steganalysis.

\subsection{Large Language Models}


Commonly, large language models (LLMs) are referred to as language models with billions of parameters that are trained on massive amounts of text. Examples of such models include GPT-3 \cite{gpt3}, PaLM \cite{chowdhery2022palm}, BLOOM \cite{bloom}, LLaMA \cite{llama} and so on. LLMs are specifically constructed using the Transformers \cite{vaswani2017attention} architecture, which stacks multiple self-attention layers in a deep network.

LLMs possess excellent natural language understanding abilities and can produce high-quality text based on given prompts. They can perform tasks such as generating code, converting formatted data into JSON files, and extracting key information from articles. Due to their powerful functionalities, smaller language models like Alpaca \cite{alpaca} and Vicuna \cite{chiangvicuna} even utilize API interfaces to directly access data from ChatGPT and GPT-4 \cite{openai2023gpt4} for training, aiming to attain similar language capabilities.
According to a previous study \cite{explore}, using ChatGPT for steganalysis tasks achieved comparable results to the second-generation steganalysis method without BERT, with just 32 examples in the prompt. Training LLMs on a dataset containing steganographic and natural texts is expected to yield improved steganalysis results.

Directly training LLMs can be costly and complex, while fine-tuning LLMs is the most frugal option. The current fine-tuning method, LoRA \cite{LoRA}, decomposes the parameter matrix of the model into two low-rank matrices, and creates a parallel LoRA branch of these matrices. The outputs of the LoRA branch are then merged with the hidden layers output of the original model. During training, the parameters of the original model are frozen, and only the two low-rank matrices in the LoRA branch are trained. This significantly reduces the parameter size, often to 1\% or lower. In most cases, the effects of LoRA fine-tuning and full fine-tuning are comparable. Considering resource efficiency and convenience, this paper adopts the LoRA method to fine-tune the LLMs.

\section{Methodology}

\subsection{Models \& Config}
We employ LoRA \cite{LoRA} to fine-tune the LLMs and directly get the detection results from the LLMs.
We utilized the Bloomz-7B1 \cite{bloomz} and the Llama-7B \cite{llama}. Both models are openly available in \url{https://huggingface.co/}. To account for our limited GPU capabilities and the trade-off between LoRA hyperparameters and training cost, we opted for LLMs with training parameters smaller than 7B. Larger models will make our GPUs out of memory. To guarantee reliable and pragmatic experimental findings, we opted for the most sizeable LLMs that our GPUs could support - the Bloomz-7B1 and Llama-7B (or Bloomz/Llama for convenience).

\begin{figure}[t]
\includegraphics[width=0.5\textwidth]{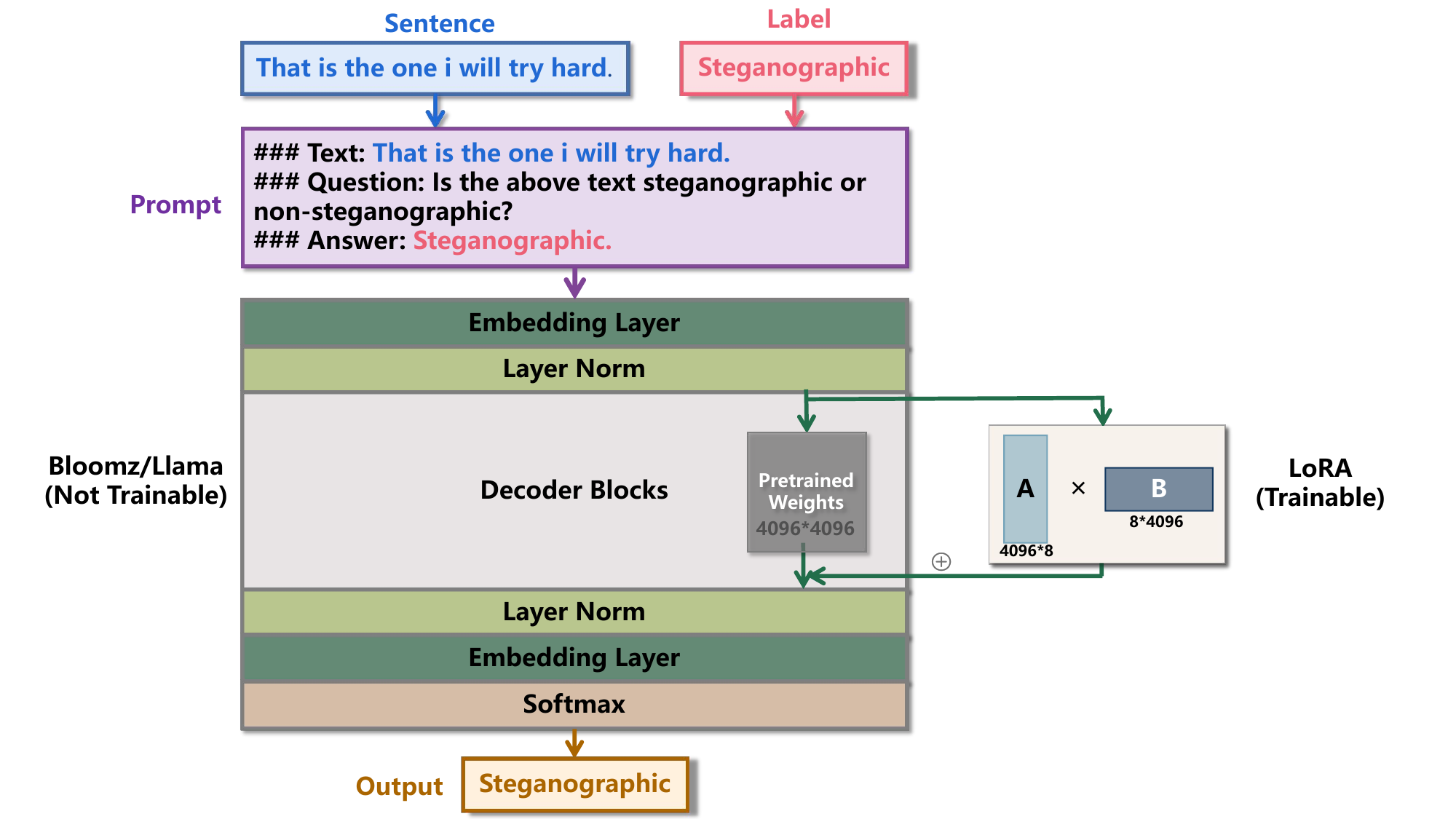}
\caption{The architecture for instruction fine-tuning in Bloomz/Llama + LoRA. Input sentences and labels are filled into a reasonable prompt template for LLM's fine-tuning. The training of parameters is limited to only the LoRA branch, while the parameters of LLMs remain constant. } 
\label{Ours model}
\end{figure}

The structure of LLMs with LoRA fine-tuning is illustrated in Fig. \ref{Ours model}. We froze all LLM parameters during the training process and exclusively trained the parameters of the LoRA branch. Thus, the NLL (Negative Log Likelihood) loss of LLM can be utilized to optimize the detection ability. Moreover, since the LLM parameters were not trained, the ability of original LLM would never decrease. This fine-tuning method guaranteed that our models could leverage the full human-like capability of original LLMs to make more precise detection.

With the LoRA fine-tune framework, trainable parameters of LLMs can be reduced to less than 0.1\%, which is much less than that of BERT. Table \ref{trainable parameters} shows the trainable parameters of our models and BERT-base. 

\begin{table}[t]
\setlength{\tabcolsep}{1pt}
\caption{Trainable Parameters}
\label{trainable parameters}
\centering
\begin{tabular}{c|ccc}
\toprule
\hline
   Model & Trainable parameters  & Total parameters & Ratio\\
\hline
   Llama-7B+LoRA  & $4,194,304$ & $6,742,609,920$ & $0.0622\%$ \\
   Bloom-7B1+LoRA & $3,932,160$ & $7,072,948,224$ & $0.0556\%$ \\
   BERT-base & $\sim 110,000,000$ & $\sim 110,000,000$ & $100\%$ \\
\hline
\bottomrule
\end{tabular}
\end{table}


\begin{table}[t]
\centering
\caption{Fixed Training Hyperparameters}\label{Fixed Parameters}
\begin{tabular}{lll}
\toprule
\hline
Model &  Bloomz & Llama \\
\hline
batch size & 4 & 4 \\
max learning rate & 1e-5 & 1e-5 \\
lora rank & 8  & 8\\
lora alpha & 16 & 16\\
lora dropout & 0.05 & 0.05\\

\hline
\bottomrule
\end{tabular}
\end{table}

Fixed hyperparameters used in training is presented in Table \ref{Fixed Parameters}. According to the report from BigScience \cite{bloomz}, the learning rate of Bloomz is set to 2e-5 during the multi-task fine-tuning phase. Hence, the learning rate for any further fine-tuning should not surpass 2e-5. The \emph{batch size} is set to 4 and the \emph{lora rank} is set to 8, which signifies the optimal configuration for executing the Bloomz-7B1 model on a single RTX3090 GPU. 
Under these conditions, the model has displayed robust performance, exhibiting the potential of LLMs in steganalysis tasks. The hyperparameter selection process of Llama is similar to that of Bloomz.
Further refinement of these hyperparameters may improve results. The previously mentioned parameter settings are intended to minimize training time and demand for GPU capacity.

\subsection{Datasets}


We utilize three datasets: Tweet, Movie, and News. We have trained copies of GPT-2 models on each dataset. Steganographic texts are generated using three different embedding methods: AC(Arithmetic Coding) \cite{ziegler2019neural}, HC(Huffman Coding) \cite{yang2018automatically}, and ADG(Adaptive Dynamic Grouping) \cite{zhang2021provably}. We randomly sampled 10,000 sentences each from natural text and steganographic-generated text to construct our datasets. These datasets are labeled as {Movie-Natural}, {Movie-AC}, {Movie-HC}, {Movie-ADG}, {Tweet-Natural}, {Tweet-AC}, {Tweet-HC}, {Tweet-ADG}, {News-Natural}, {News-AC}, {News-HC}, and {News-ADG}. The first part of each label represents the data source, and the second part represents the steganographic encoding algorithm. Table \ref{Datasets} presents the statistical features of each dataset.

\begin{table}[t]
\centering
\setlength{\tabcolsep}{4pt}
\setlength{\belowrulesep}{0pt}
\caption{Average perplexity and tokens of the datasets}\label{Datasets}
\begin{tabular}{c|cccc}
\toprule
\midrule
        Features & Tweet-Natural & Tweet-AC & Tweet-HC & Tweet-ADG \\ \hline
        Tokens & 10.66  & 10.78  & 9.44  & 9.74  \\ 
        PPL & 946.16  & 1016.45  & 635.35  & 1099.58  \\ \hline\hline
        Features & Movie-Natural & Movie-AC & Movie-HC & Movie-ADG \\ \hline
        Tokens & 25.63  & 26.48  & 24.00  & 27.80  \\ 
        PPL &  254.33 & 261.51  & 119.20  & 256.12  \\ \hline\hline
        Features & News-Natural & News-AC & News-HC & News-ADG \\ \hline
        Tokens & 23.10  & 26.10  & 24.42  & 25.93  \\ \
        PPL & 253.65  & 242.22  & 109.87  & 248.75  \\
        \hline
        \bottomrule
\end{tabular}
\end{table}

\begin{figure*}[ht]
\subfigure[Distribution of natural text]{
\includegraphics[width=0.22\textwidth]{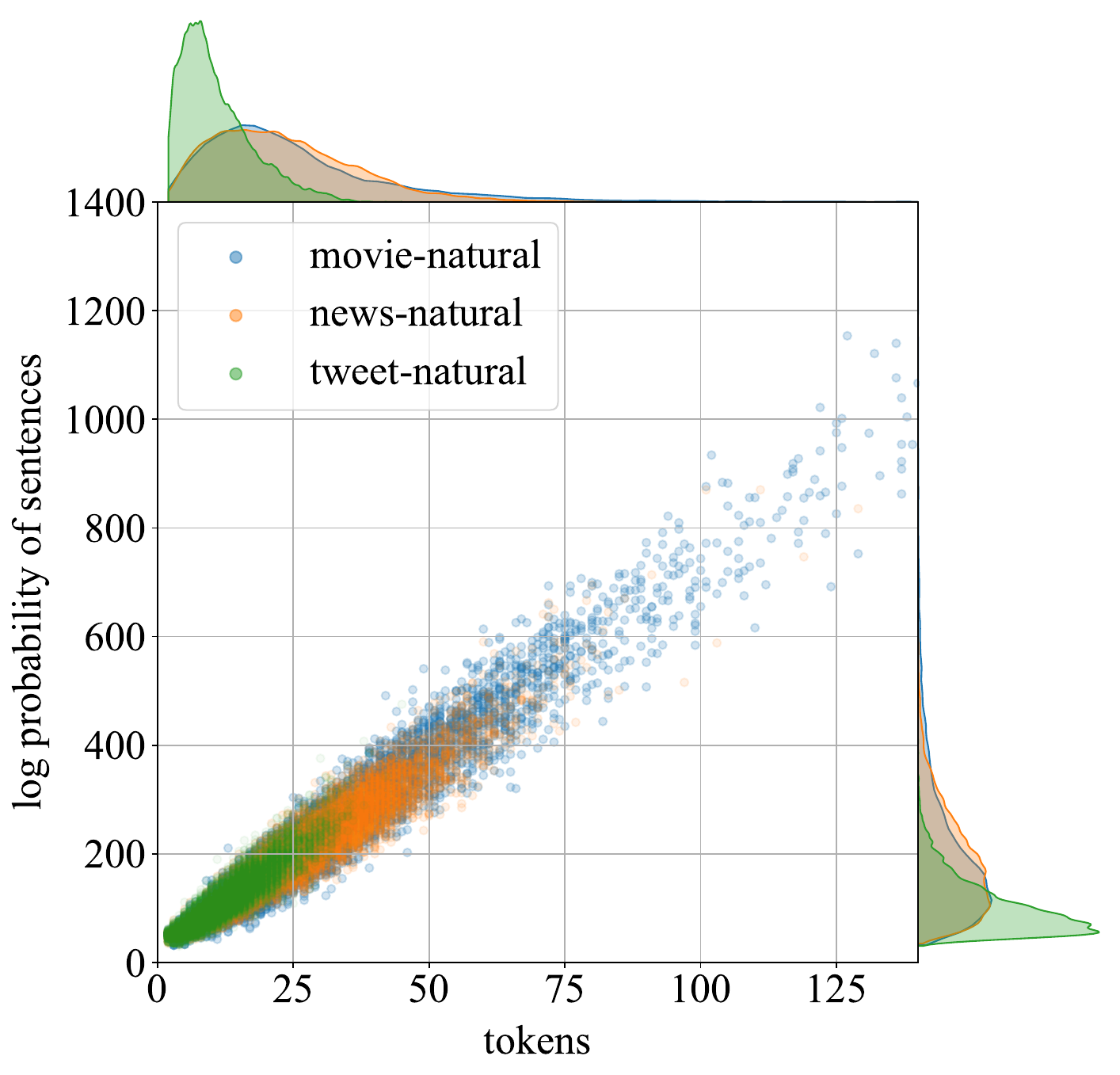}
\label{Natural human distributions}}
\hspace{0.01cm}
\subfigure[Distribution of Movie stegos]{
\includegraphics[width=0.22\textwidth]{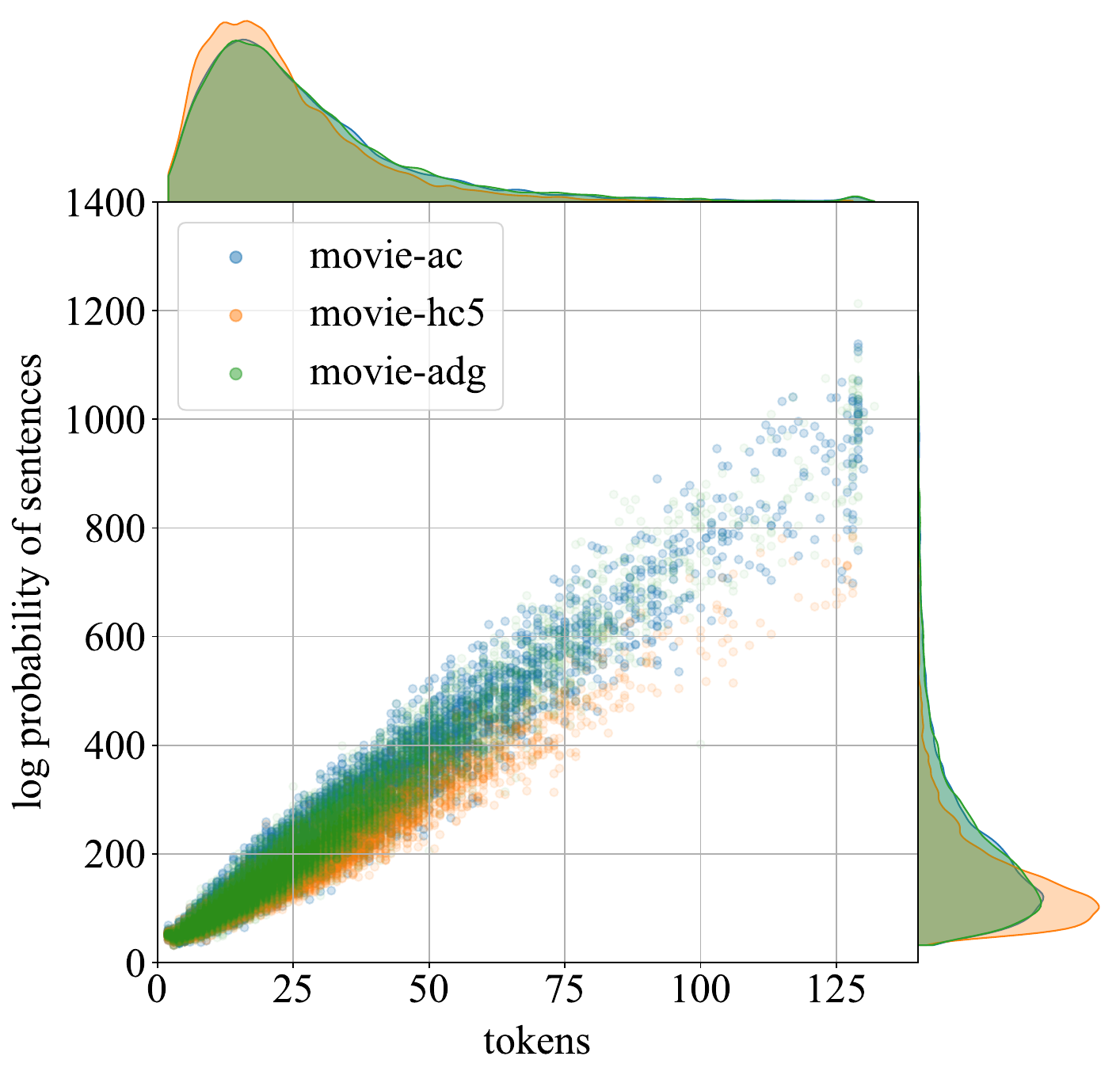}
\label{Movie-AC/HC/ADG distributions}}
\hspace{0.01cm}
\subfigure[Distribution of News stegos]{
\includegraphics[width=0.22\textwidth]{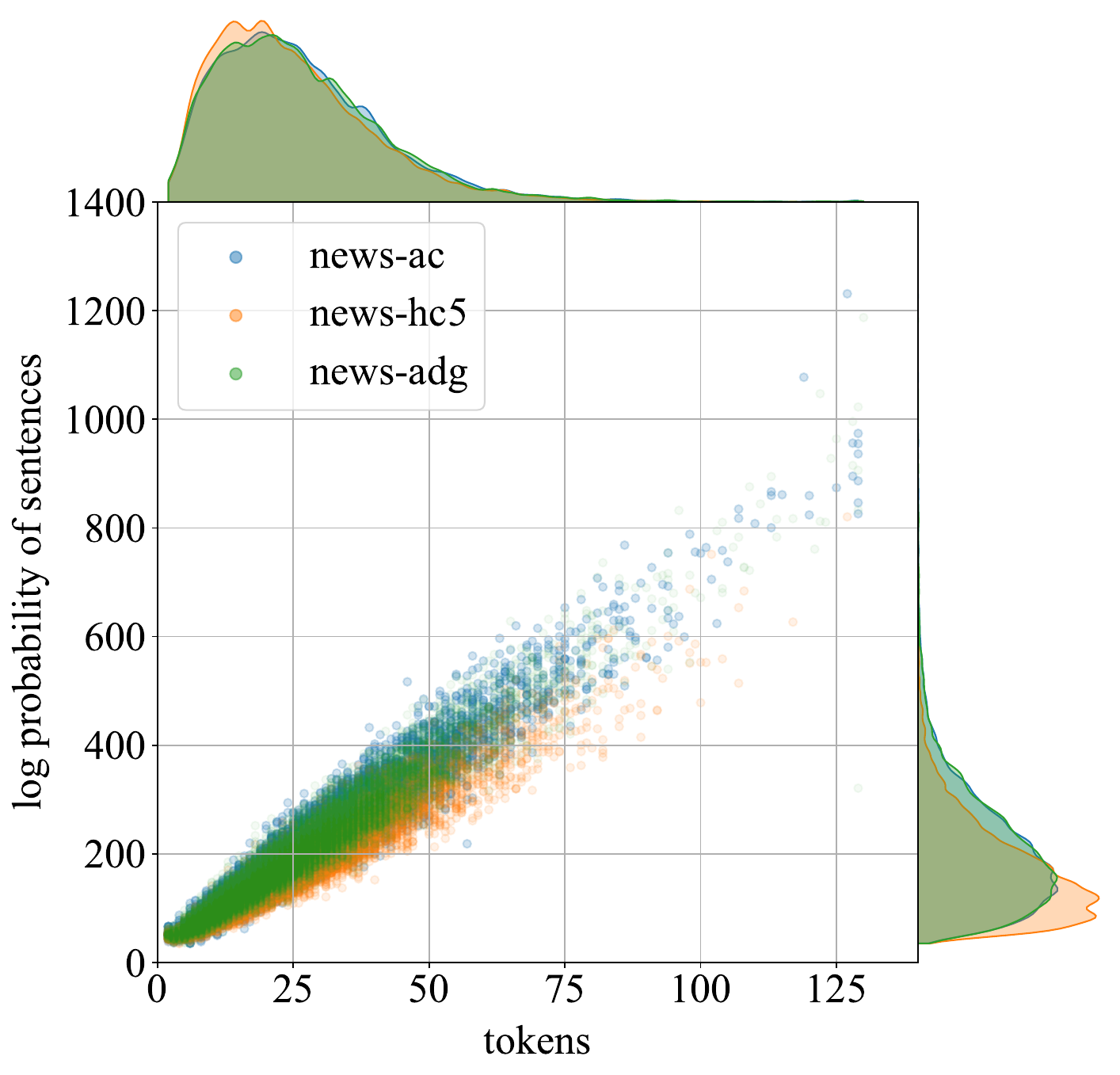} 
\label{News-AC/HC/ADG distributions}}
\hspace{0.01cm}
\subfigure[Distribution of Tweet stegos]{
\includegraphics[width=0.22\textwidth]{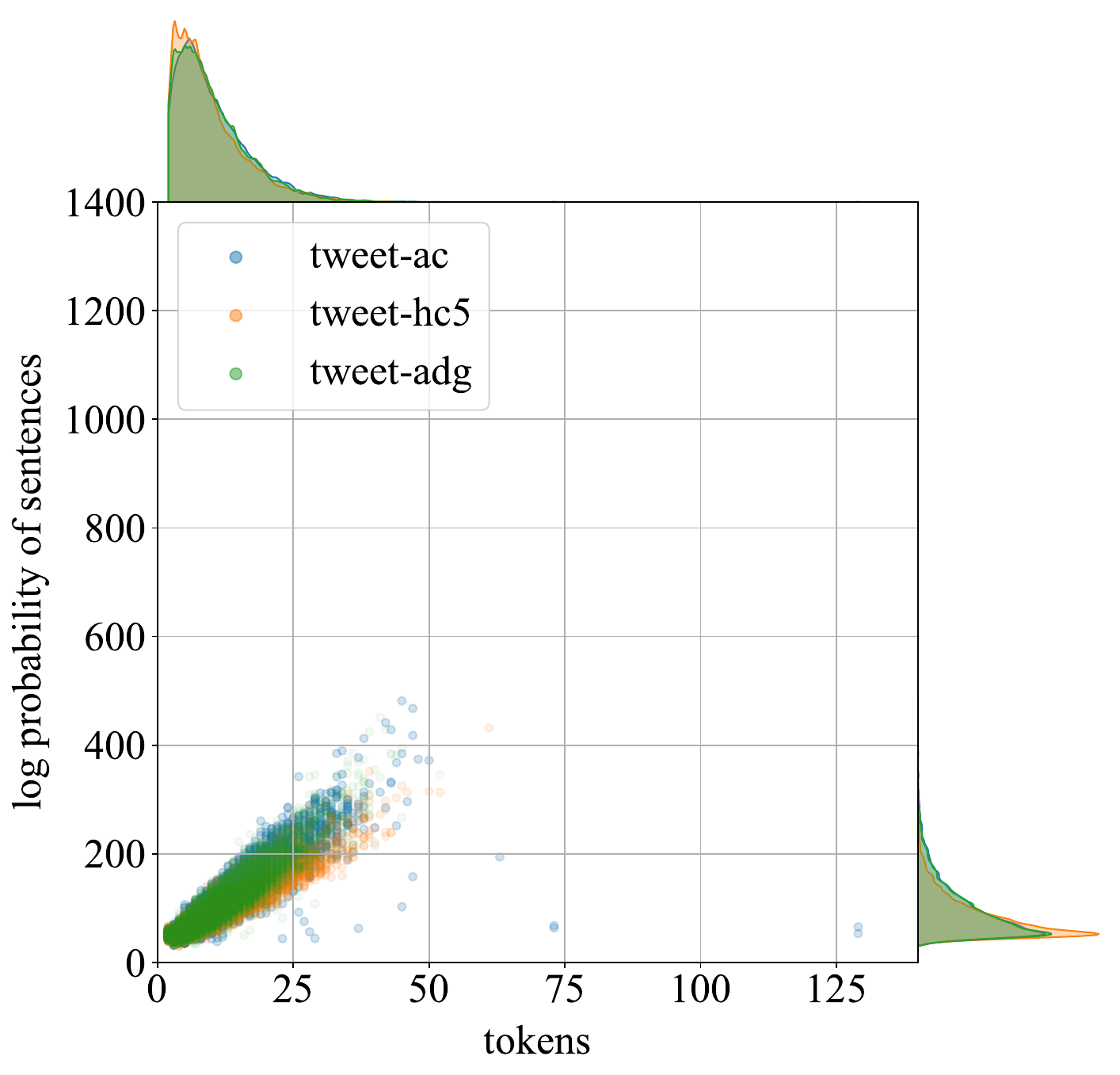} 
\label{Tweet-AC/HC/ADG distributions}}
\\
\subfigure[Normalized distributions of natural text]{
\includegraphics[width=0.22\textwidth]{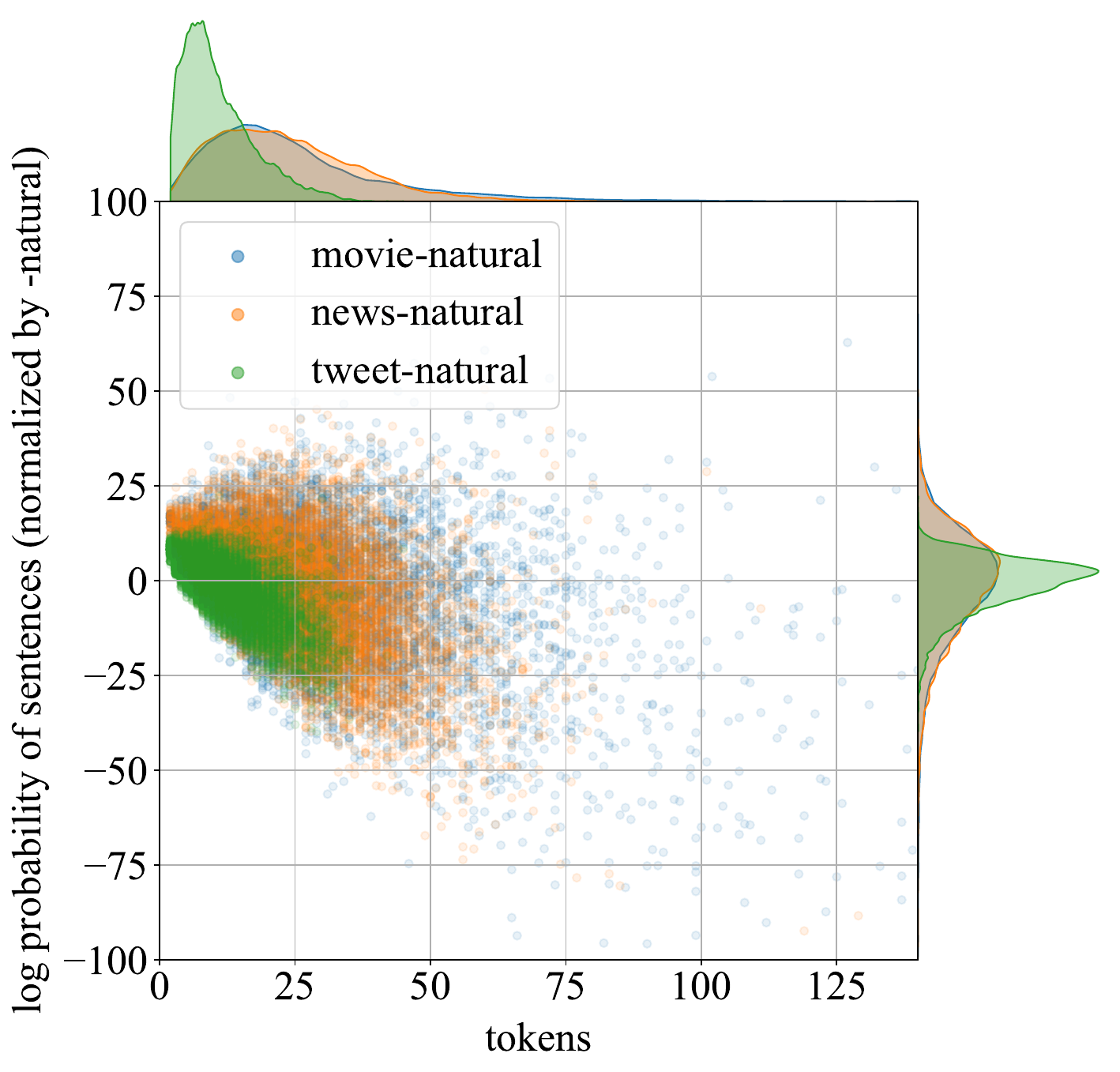}
\label{Natural human normalized distributions}}
\hspace{0.01cm}
\subfigure[Normalized distributions of Movie stegos]{
\includegraphics[width=0.22\textwidth]{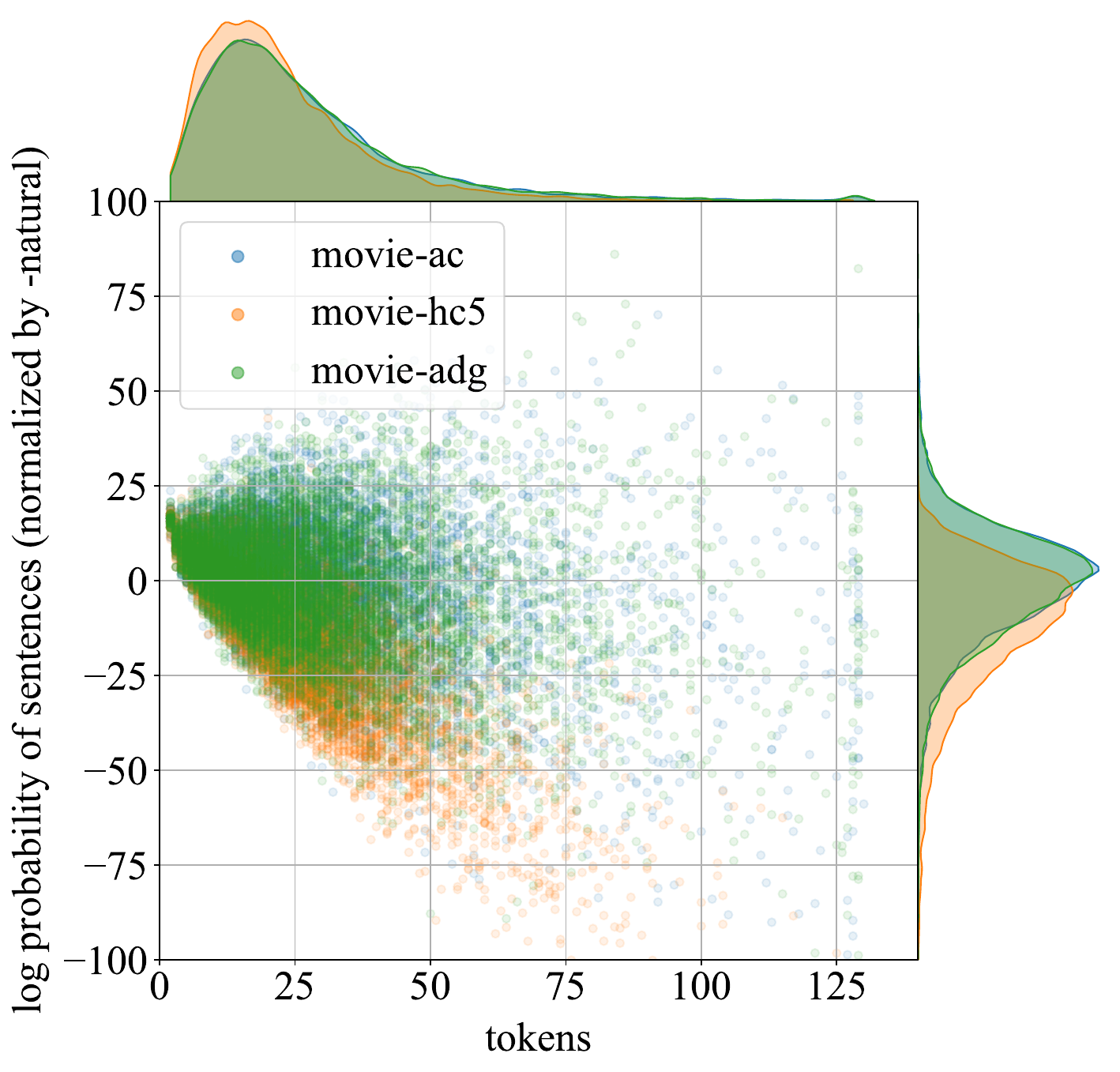}
\label{Movie-AC/HC/ADG normalized distributions}}
\hspace{0.01cm}
\subfigure[Normalized distributions of News stegos]{
\includegraphics[width=0.22\textwidth]{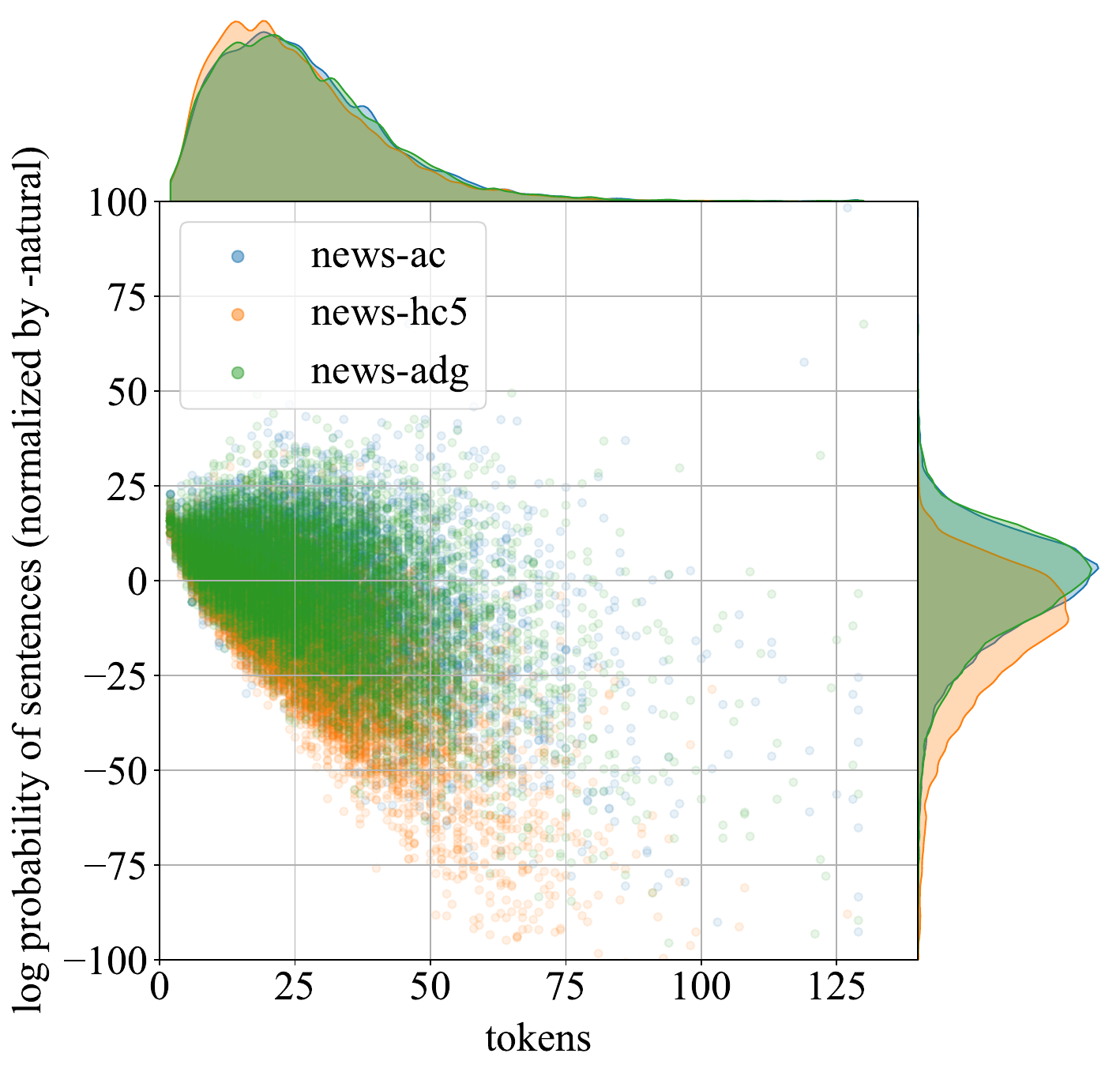} 
\label{News-AC/HC/ADG normalized distributions}}
\hspace{0.01cm}
\subfigure[Normalized distributions of Tweet stegos]{
\includegraphics[width=0.22\textwidth]{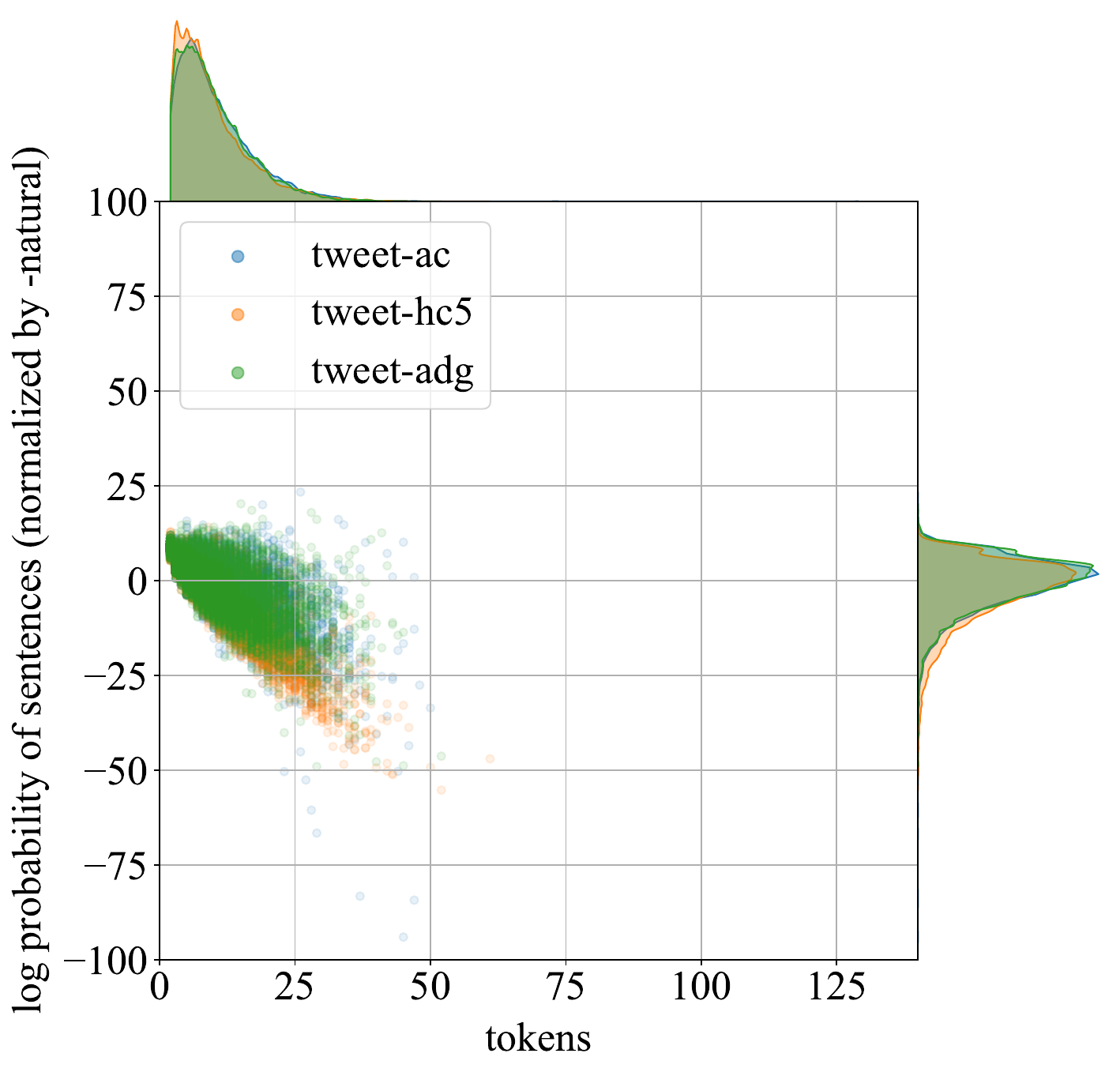} 
\label{Tweet-AC/HC/ADG normalized distributions}}
\caption{Distributions of text, estimated by log probability of sentences and the number of tokens in sentences. Each figure consists of the main scatter plot of joint distribution and lines of marginal distributions at top and right. (a) represents the distributions of 3 natural datasets, and (b) (c) (d) denote the distributions of AC/HC/ADG stegos within these datasets, respectively. The second row illustrates the normalized distribution, estimated by normalized log probability of sentences and the number of tokens in sentences, corresponding to the first row.} 
\label{Transferability of all models}
\end{figure*}


We modeled the distributions of datasets in terms of sentence length and perplexity.
Sentence length is the number of tokens that the text breaks down under the tokenizer of the model. 
And perplexity (PPL) is a widely employed metric to assess the fluency of text, whereas entropy represents a statistical measure of the information content. The formulas for PPL and entropy can be expressed as follows.
\begin{equation}
\text{PPL} = \exp\left(-\frac{1}{N}\sum_{i=1}^{N}\log p(x_i|\textbf{x}_{1:i-1})\right)
\end{equation}

where $N$ represents the total number of tokens, and $p(x_i)$ denotes the conditional probability assigned to each token $x_i$.



It is evident that the statistical characteristics of the {Tweet} dataset differ significantly from those of the {Movie} and {News} datasets. The average length of the {Tweet} is less than half of the average length observed in the {Movie} and {News} datasets, whereas the PPL is nearly 3 times higher. This suggests that the texts in {Tweet} are disorganized and confused. In contrast, the statistical characteristics of {Movie} and {News} exhibit relatively similar patterns.

The three steganographic encoding algorithms display distinct characteristics. The steganographic texts created by HC possess the lowest PPL, surpassing even natural texts. Conversely, the steganographic texts produced by AC closely emulate the statistical properties of natural texts. In the case of ADG-generated texts, they demonstrate a similar pattern to AC-generated texts but with a higher perplexity.

It should be noted that PPL is not considered the definitive criterion for assessing the quality of text generation. Upon manual evaluation, both the steganographic texts produced by AC and ADG exhibit considerable similarity. PPL merely offers a concise overview of the distribution of text in the dataset and is inadequate for discerning between high and low-quality text.

To more accurately characterize the distribution of the dataset, we calculated the log probability of sentences, then used log probability as the x-axis and sentence length (counted by tokens) as the y-axis to model the distribution of text. The distribution of log probability and sentence length can be viewed as a simplified 2-dimensional text probability distribution. The log probability can be calculated as Equation \ref{logprob}:
\begin{equation}\label{logprob}
    log\,probability = -\sum_{i = 1}^{N} \log p(x_i|\textbf{x}_{1:i-1}),
\end{equation}
        where $N$ denotes the length of the sentence.

As illustrated in Fig. \ref{Natural human distributions}, \ref{Movie-AC/HC/ADG distributions}, \ref{News-AC/HC/ADG distributions}, and \ref{Tweet-AC/HC/ADG distributions}, the distribution of the Tweet dataset is primarily focused on short sentences with a relatively high log probability. Although the distributions of the Movie and News datasets are similar, the main discrepancy lies in the fact that the long-tailed distribution of the Movie dataset constitutes a larger proportion.

To elucidate the influence of steganographic algorithms on the distribution of text, we standardized the logarithmic probability of AC/HC/ADG using the mean and standard deviation of natural text.
Based on Fig. \ref{Natural human normalized distributions},\ref{Movie-AC/HC/ADG normalized distributions},\ref{News-AC/HC/ADG normalized distributions},\ref{Tweet-AC/HC/ADG normalized distributions}, it is clear that the distribution of stegos produced by HC is lower than that of the other distributions. The distribution of edges in HC stegos reveals that HC tends to produce text that significantly deviates from natural human writing patterns. This is due to its peak position of log probability is less than 0, resulting in a noticeable distinguishability of HC-generated stegos. Conversely, AC and ADG stegos display a nearly identical distribution; however, the AC distribution appears to be more focused than that of ADG.

\begin{table*}[ht]
\setlength{\tabcolsep}{3pt}
  \centering
  \caption{Results on Movie, News and Tweet. ``w/o FT" represents the zero-shot output without fine-tuning. The baseline models \cite{TS-CSW,TS-RNN,yang2019fast,zou2020high,niu2019hybrid,yang2020linguistic,EILGF} and our models are trained for 10 epochs on each dataset.}
    \begin{tabular}{l|cccc|cccc|cccc}
    \toprule
    \hline
    \multirow{2}{*}{Models} & \multicolumn{4}{c|}{Movie-AC}  & \multicolumn{4}{c|}{Movie-HC}  & \multicolumn{4}{c}{Movie-ADG} \\
    & {Accuracy} & {F1 score} & {Precision} & {Recall} & {Accuracy} & {F1 score} & {Precision} & {Recall} & {Accuracy} & {F1 score} & {Precision} & {Recall}\\\hline
    TS-CSW \cite{TS-CSW} & 68.17  & 67.92  & 71.27  & 64.87  & 79.05  & 80.23  & 75.30  & 85.86  & 71.80  & 68.81  & 76.09  & 62.80  \\
    TS-RNN \cite{TS-RNN} & 72.20  & 69.31  & 76.44  & 63.40  & 81.15  & 81.75  & 78.52  & 85.26  & 72.03  & 68.93  & 76.60  & 62.65  \\
    TS-FCN \cite{yang2019fast}& 73.35  & 72.01  & 75.04  & 69.21  & 81.15  & 81.89  & 78.10  & 86.07  & 72.80  & 70.45  & 76.25  & 65.47  \\
    LSTMATT \cite{zou2020high}& 74.40  & 72.57  & 77.30  & 68.40  & 82.00  & 82.95  & 78.11  & 88.44  & 66.88  & 65.32  & 67.82  & 63.00  \\
    RbiLSTMC \cite{niu2019hybrid}& 70.88  & 67.72  & 75.06  & 61.68  & 78.90  & 80.10  & 75.14  & 85.76  & 72.03  & 68.96  & 76.54  & 62.75  \\
    BiLSTMDENSE \cite{yang2020linguistic}& 69.30  & 59.55  & 87.34  & 45.18  & 80.65  & 81.70  & 76.83  & 87.23  & 73.05  & 70.19  & 77.61  & 64.06  \\
    EILGF \cite{EILGF}& 69.42  & 67.83  & 72.88  & 63.44  & 76.67  & 75.57  & 74.40  & 76.77  & 66.50  & 64.61  & 68.22  & 61.37  \\ \hline
    ChatGPT (32 shots) \cite{explore} & 66.60 & 73.58 & 60.86 & 93.00 & 65.80 & 68.04 & 64.02 & 72.60 & 70.00 & 72.58 & 66.84 & 79.40\\
    Bloomz w/o FT & 43.65 & 51.85 & 45.48 & 60.30 & 43.70 & 51.91 & 45.52 & 60.40 & 43.30 & 51.40 & 41.85 & 59.60\\
    Llama w/o FT & 52.50 & 60.87 & 51.79 & 73.80 & 53.70 & 62.23 & 52.59 & 76.20 & 53.25 & 61.72 & 52.29 & 75.30\\\hline
    Bloomz w/ FT& 86.50  & 86.53  & 86.35  & 86.70  & 88.85  & 89.14  & 86.86  & 91.55  & 86.05  & 86.05  & 86.05  & 86.05  \\
    Llama w/ FT& \textbf{89.35}  & \textbf{89.55}  & 87.87  & 91.30  & \textbf{90.95}  & \textbf{91.01}  & 90.42  & 91.60  & \textbf{89.15}  & \textbf{89.08}  & 89.67  & 88.50  \\\hline
    \hline
    \multirow{2}{*}{Models} & \multicolumn{4}{c|}{News-AC}  & \multicolumn{4}{c|}{News-HC}  & \multicolumn{4}{c}{News-ADG} \\
    & {Accuracy} & {F1 score} & {Precision} & {Recall} & {Accuracy} & {F1 score} & {Precision} & {Recall} & {Accuracy} & {F1 score} & {Precision} & {Recall}\\\hline
    TS-CSW \cite{TS-CSW}& 70.25  & 68.28  & 72.33  & 64.66  & 77.90  & 79.64  & 73.23  & 87.28  & 70.00  & 67.62  & 72.64  & 63.25  \\
    TS-RNN \cite{TS-RNN}& 70.88  & 67.70  & 61.64  & 61.64  & 79.08  & 80.16  & 75.56  & 85.36  & 70.15  & 66.52  & 74.82  & 59.87  \\
    TS-FCN \cite{yang2019fast}& 71.02  & 69.03  & 73.32  & 65.22  & 79.18  & 80.93  & 74.06  & 89.20  & 71.50  & 67.89  & 76.80  & 60.83  \\
    LSTMATT \cite{zou2020high}& 67.77  & 65.96  & 69.16  & 63.05  & 74.55  & 74.95  & 73.11  & 76.88  & 71.95  & 69.14  & 75.95  & 63.45  \\
    RbiLSTMC \cite{niu2019hybrid}& 70.30  & 66.80  & 74.83  & 60.32  & 79.30  & 80.26  & 76.05  & 84.96  & 69.60  & 66.01  & 73.95  & 59.61  \\
    BiLSTMDENSE \cite{yang2020linguistic}& 68.75  & 63.41  & 75.47  & 54.67  & 78.10  & 79.97  & 73.08  & 88.28  & 69.88  & 66.92  & 73.34  & 61.53  \\
    EILGF \cite{EILGF}& 61.83  & 63.53  & 62.74  & 64.35  & 73.58  & 73.39  & 75.61  & 71.29  & 61.17  & 61.42  & 62.88  & 60.03  \\\hline
    ChatGPT (32 shots) \cite{explore}& 66.30 & 71.37 & 62.04 & 84.00 & 59.30 & 56.28 & 60.79 & 52.40 & 67.30 & 66.39 & 68.51 & 64.40\\
    Bloomz w/o FT & 47.40 & 45.05 & 47.56 & 42.80 & 49.45 & 48.33 & 49.84 & 46.90 & 48.55 & 46.91 & 48.86 & 45.10 \\
    Llama w/o FT & 49.10 & 58.71 & 49.42 & 72.30 & 51.15 & 61.02 & 50.80 & 76.40 & 49.35 & 59.00 & 49.59 & 72.80\\\hline
    Bloomz w/ FT & 82.85  & 82.80  & 83.05  & 82.55  & 88.33  & 88.45  & 87.48  & 89.45  & 83.53  & 83.92  & 81.94  & 86.00  \\
    Llama w/ FT & \textbf{87.55}  & \textbf{87.74}  & 86.42  & 89.10  & \textbf{89.95}  & \textbf{89.94}  & 89.99  & 89.90  & \textbf{88.45}  & \textbf{88.59}  & 87.51  & 89.70  \\\hline
    \hline
    \multirow{2}{*}{Models} & \multicolumn{4}{c|}{Tweet-AC}  & \multicolumn{4}{c|}{Tweet-HC}  & \multicolumn{4}{c}{Tweet-ADG} \\
    & {Accuracy} & {F1 score} & {Precision} & {Recall} & {Accuracy} & {F1 score} & {Precision} & {Recall} & {Accuracy} & {F1 score} & {Precision} & {Recall}\\\hline
    TS-CSW \cite{TS-CSW}& 66.02  & 64.64  & 66.70  & 62.69  & 71.30  & 73.28  & 67.99  & 79.45  & 64.58  & 63.93  & 64.48  & 63.40  \\
    TS-RNN \cite{TS-RNN}& 66.72  & 33.19  & 33.32  & 65.77  & 72.15  & 73.50  & 69.50  & 77.99  & 65.22  & 64.68  & 65.07  & 64.31  \\
    TS-FCN \cite{yang2019fast}& 68.25  & 67.43  & 68.52  & 66.38  & 71.23  & 73.71  & 67.31  & 81.47  & 66.63  & 66.52  & 66.10  & 66.93  \\
    LSTMATT \cite{zou2020high}& 63.20  & 60.79  & 64.35  & 57.60  & 68.38  & 70.21  & 65.80  & 75.26  & 64.67  & 63.41  & 76.44  & 70.62  \\
    RbiLSTMC \cite{niu2019hybrid}& 67.35  & 67.05  & 67.02  & 67.09  & 71.73  & 73.96  & 67.99  & 81.07  & 66.10  & 65.16  & 65.35  & 64.01  \\
    BiLSTMDENSE \cite{yang2020linguistic}& 66.70  & 67.11  & 65.68  & 68.60  & 70.63  & 73.13  & 66.85  & 80.72  & 65.63  & 65.15  & 65.43  & 64.87  \\
    EILGF \cite{EILGF}& 68.00  & 67.07  & 69.20  & 65.06  & 69.75  & 64.52  & 74.66  & 56.80  & 67.17  & 67.60  & 66.40  & 68.84  \\\hline
    ChatGPT (32 shots) \cite{explore}& 61.10 & 63.75 & 59.69 & 68.40 & 48.70 & 57.24 & 49.42 & 68.00 & 56.20 & 65.07 & 54.11 & 81.60 \\
    Bloomz w/o FT & 54.95 & 63.61 & 53.56 & 78.30 & 56.30 & 65.09 & 54.40 & 81.00 & 54.10 & 62.66 & 53.01 & 76.60\\
    Llama w/o FT & 49.90 & 65.66 & 49.95 & 95.80 & 50.85 & 66.53 & 50.44 & 97.70 & 49.60 & 65.38 & 49.79 & 95.20\\\hline
    Bloomz w/ FT & 76.78  & 76.84  & 76.63  & 77.05  & 79.73  & 80.57  & 77.33  & 84.10  & 75.08  & 75.13  & 74.96  & 75.30  \\
    Llama w/ FT & \textbf{79.30}  & \textbf{79.84}  & 77.80  & 82.00  & \textbf{82.35}  & \textbf{82.62}  & 81.38  & 83.90  & \textbf{78.40}  & \textbf{79.13}  & 76.54  & 81.90  \\\hline
    \bottomrule
    \end{tabular}
    \label{tweet}%
\end{table*}%

\subsection{Evaluation Metrics}
We denote the output results of the model as TS (True Steganographic), FS (False Steganographic), US (Unknown Steganographic), TN (True Non-Steganographic), FN (False Non-Steganographic), and UN (Unknown Non-Steganographic), where US and UN represent samples that do not match the output template.  The occurrence of US and UN is not unusual in \cite{explore}, but in our models they only appear when the models are not fine-tuned. We primarily use accuracy and F1 score as evaluation metrics. Since the existence of US and UN samples, redefining the calculation of metrics is necessary. The accuracy is redefined as follows:

\begin{equation}
    Acc = \frac{TS+TN}{TS+FS+US+TN+FN+UN}.
    \end{equation}


Due to the equitable sampling of steganographic and natural texts in our dataset, and taking into account the balance between recall and precision, we utilize the F1 score with $\beta = 1$ as the principal evaluation metric for steganalysis. The F1 score is obtained as follows:

\begin{equation}
    Precision = \frac{TS}{TS+FS+US},
    \end{equation}
\begin{equation}
    Recall = \frac{TS}{TS+FN},
    \end{equation}
\begin{equation}
    F1 = (1+\beta^2) \cdot \frac{Precision \cdot Recall}{\beta^2 \cdot Precision + Recall}, \beta = 1.
    \end{equation}
    

The expression for F1 can be obtained by substituting the formulas for Precision and Recall, as follows:
\begin{equation}
    F1 = \frac{2 \cdot TS}{2 \cdot TS + FN + FS + US}.
    \end{equation}

\subsection{Prompt Design}
\label{main prompt}
We primarily employ the following templates to construct the instruction fine-tuning dataset:

\textit{\#\#\# Text:\{input sentence\}}

\textit{\#\#\# Question: Is the above text steganographic or non-steganographic?}

\textit{\#\#\# Answer: \{input label\}}










The effects of the prompts and their impact on steganalysis will be thoroughly discussed in Section \ref{Effectiveness of different prompts}.

\section{Experiments and Analysis}

\begin{figure*}[ht]
\centering
\includegraphics[width=1\textwidth]{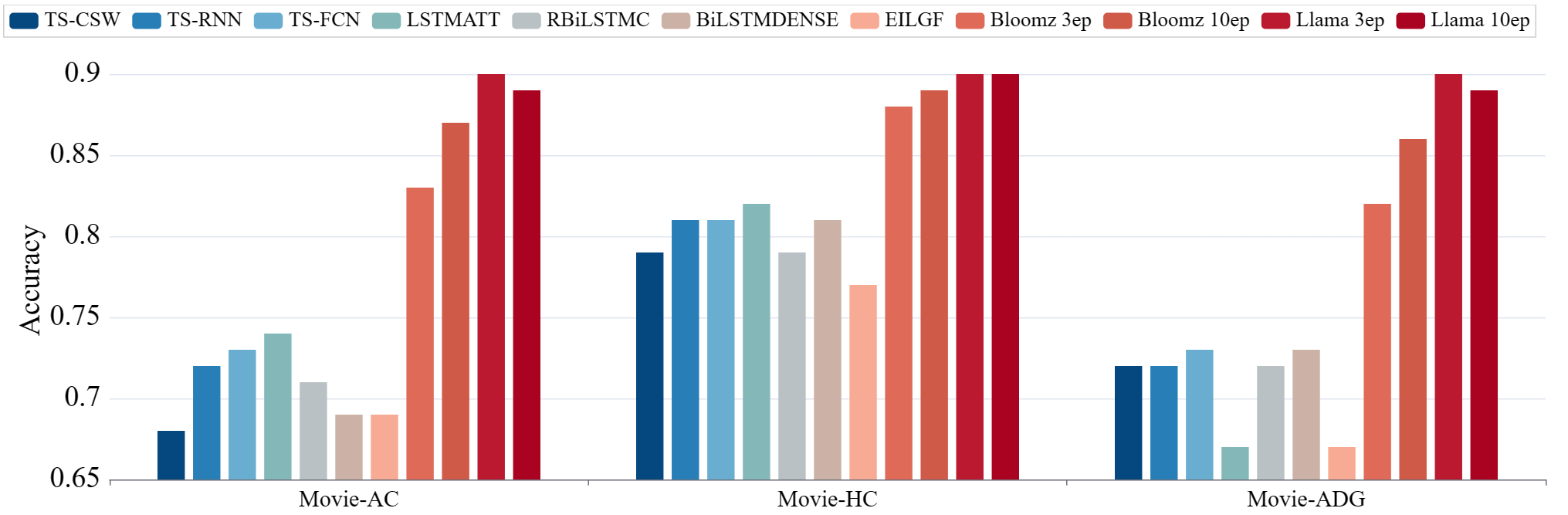} \label{Common Compare Movie}
\caption{Detection accuracy of Bloomz/Llama training with 3/10 epochs and baseline methods training with 10 epochs on {Movie} datasets. } 
\label{common compare}
\end{figure*}
To verify the effectiveness and generality of our method, we first tested in domain-specific steganalysis task and compared the results with current advanced steganalysis methods. Then we analyzed the error samples of this task and tried to find some clues for training a general steganalysis model. After that, we tested our methods in a more practical and challenging scenario of domain-agnostic steganalysis. To figure out the critical factors that affected the performance of our method, we investigated the detection accuracy on sentences of various length and different training prompts. Finally, we trained a general steganalysis model that was effective in diverse tasks.

The baseline steganalysis methods used for comparison are as follows: TS-FCN \cite{yang2019fast} utilizes a single-layer fully connected network to discern statistical correlation at the word level. TS-RNN \cite{TS-RNN} employs a two-layer bidirectional RNN to decide on temporal connections between words. TS-CSW \cite{TS-CSW} incorporates a multi-scale kernel CNN in its discriminator, enabling the recognition of statistical features of various sizes. RBiLSTMC \cite{niu2019hybrid} combines the advantages of Bi-LSTM and CNN to enhance detection accuracy. LSTMATT \cite{zou2020high} utilizes the attention mechanism. BiLSTMDENSE \cite{yang2020linguistic} improves low-level features through dense connections and feature pyramids. EILGF \cite{EILGF} extracts and integrates local and global features.

These methods encompass classical and efficient steganalysis models, as well as the latest and state-of-the-art steganalysis models.

\subsection{Domain-Specific Steganalysis}

We conduct domain-specific steganalysis experiments within a dataset of steganographic texts and a corresponding dataset of natural texts. These 2 datasets are divided into training, validation, and test sets in a ratio of 3:1:1. This ratio is consistently maintained for all subsequent experiments.
The Bloomz/Llama and the baseline models were compared on all datasets, as shown in Table \ref{tweet}. The result show that without fine-tuning the LLMs are failed to demonstrate the capability of steganalysis. After fine-tuning the detection accuracy and F1 score of LLMs outperform all baseline methods on all datasets.
Fig. \ref{common compare} shows a comparison of the Bloomz training with 3/10 epochs, Llama training with 3/10 epochs, and the baseline methods training with 10 epochs. Results show that both Bloomz and Llama consistently outperform the baseline methods in all scenarios, with significant improvements in AC and ADG. It is imperative to clarify that the validation set utilized in our experiments was only used for comparison with baseline methods and not incorporated by our models.

\begin{figure}[t]

\includegraphics[width=0.5\textwidth]{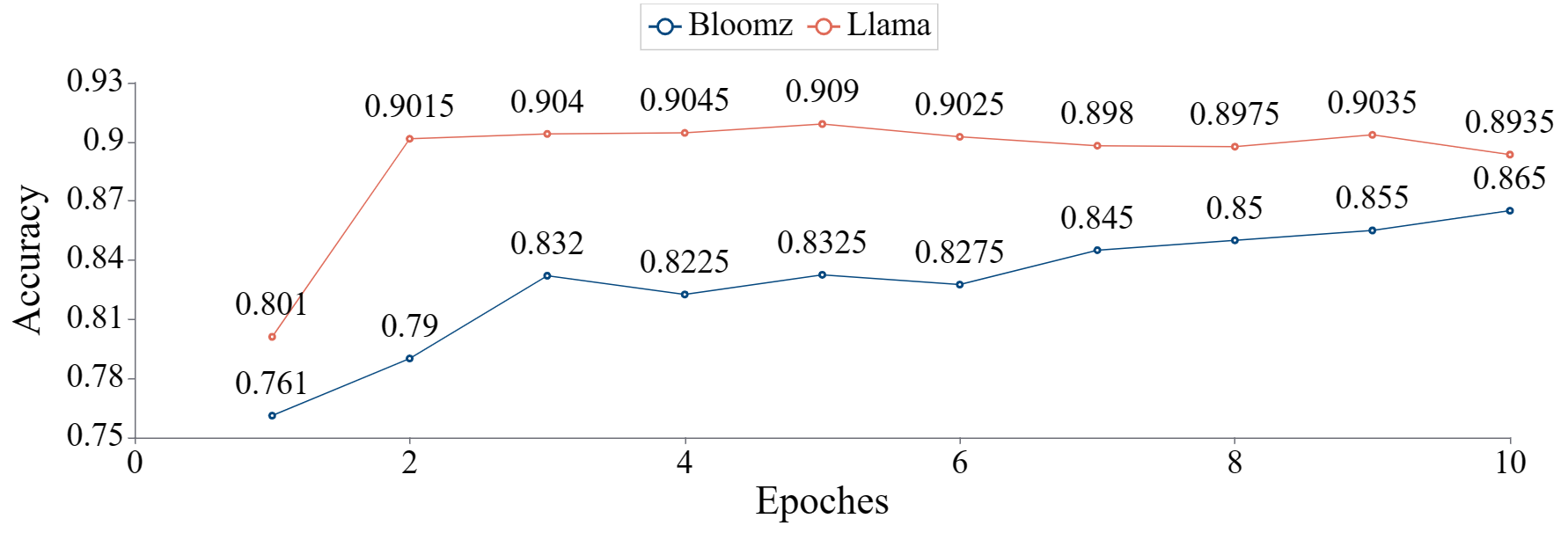} 

\caption{Detection accuracy of Bloomz/Llama training with 1-10 epochs on Movie-AC dataset.}
\label{Different Epoches}
\end{figure}
The results obtained from the Movie-ADG dataset reveal that Bloomz yields a superior accuracy rate of approximately 13\% and an F1 score that is around 16\% higher than the optimal baseline model. Moreover, the more effective Llama poses an accuracy rate that is roughly 16\% higher and an F1 score of approximately 19\%, with a detection success rate of nearly 90\%. Furthermore, detection accuracy on the three steganographic encoding algorithms achieved at least 90\% in the Movie dataset, while not appearing the discrepancy seen in the baseline methods, where detection accuracy on HC-generated texts was notably higher than that of AC and ADG.

\begin{figure*}[ht!]
\subfigure[Distribution of error samples in Movie-AC]{
\includegraphics[width=0.31\textwidth]{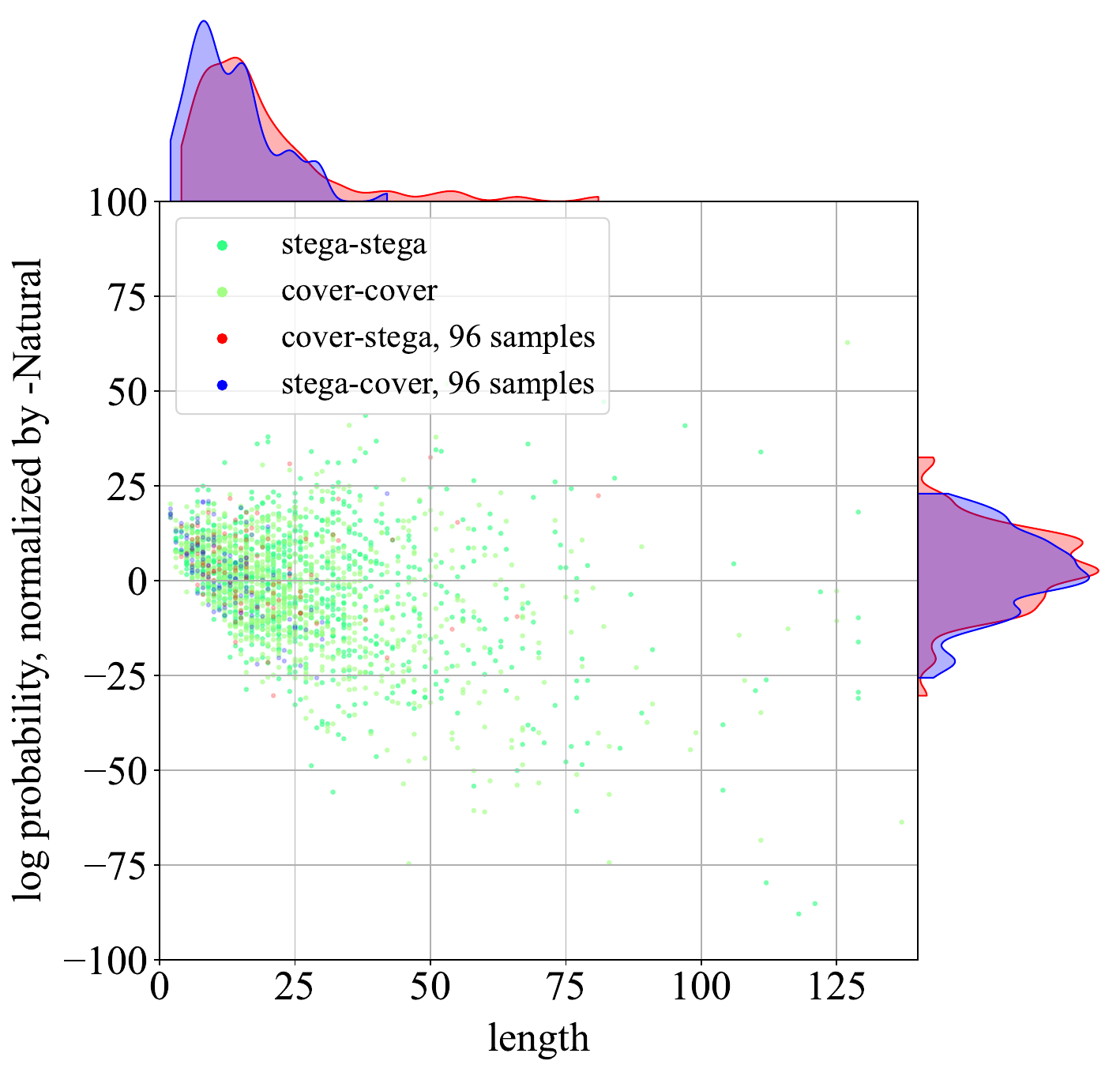}
\label{Movie-AC error distribution}}
\subfigure[Distribution of error samples in Movie-HC]{
\includegraphics[width=0.31\textwidth]{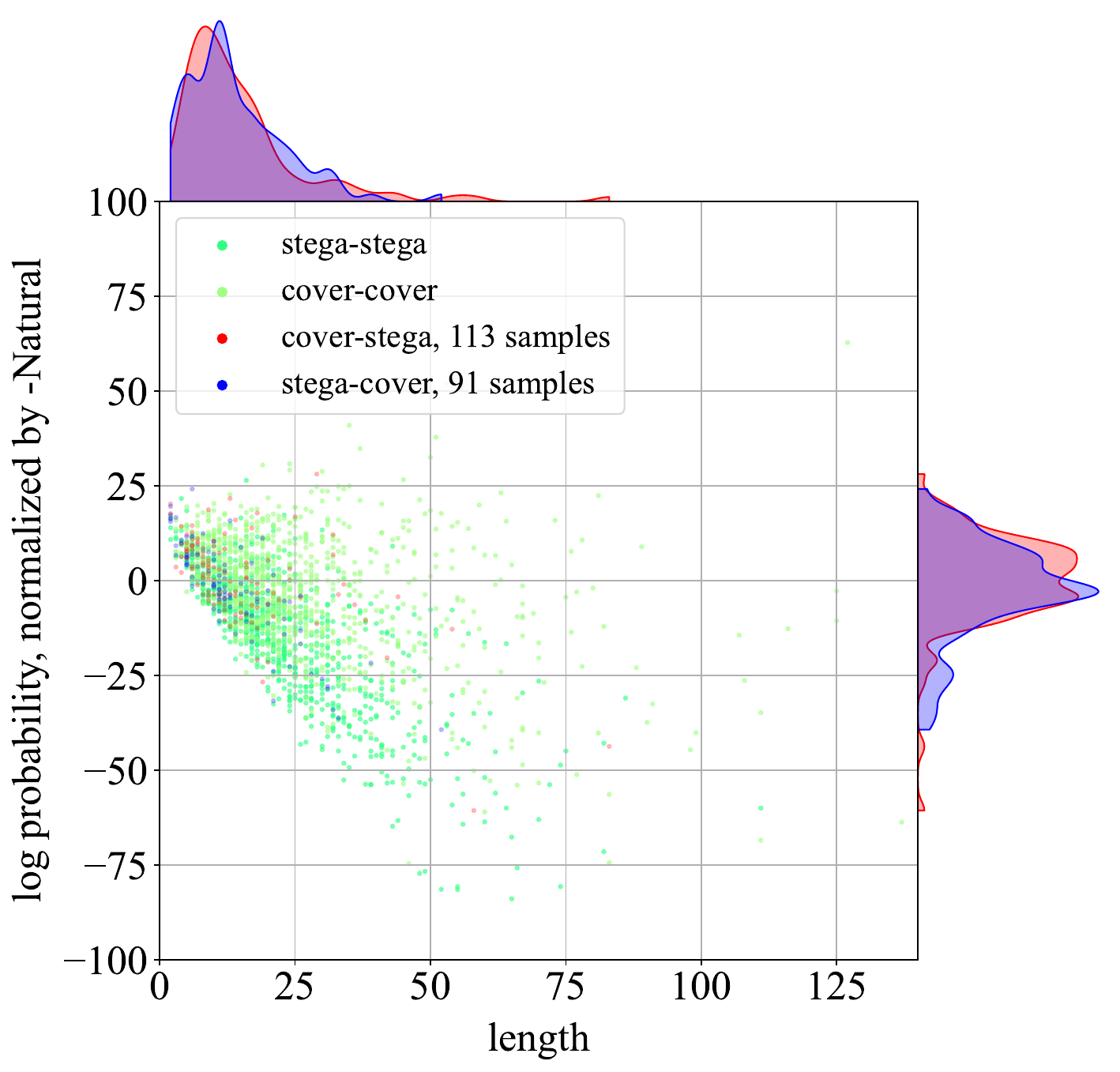}
\label{Movie-HC error distribution}}
\subfigure[Distribution of error samples in Movie-ADG]{
\includegraphics[width=0.31\textwidth]{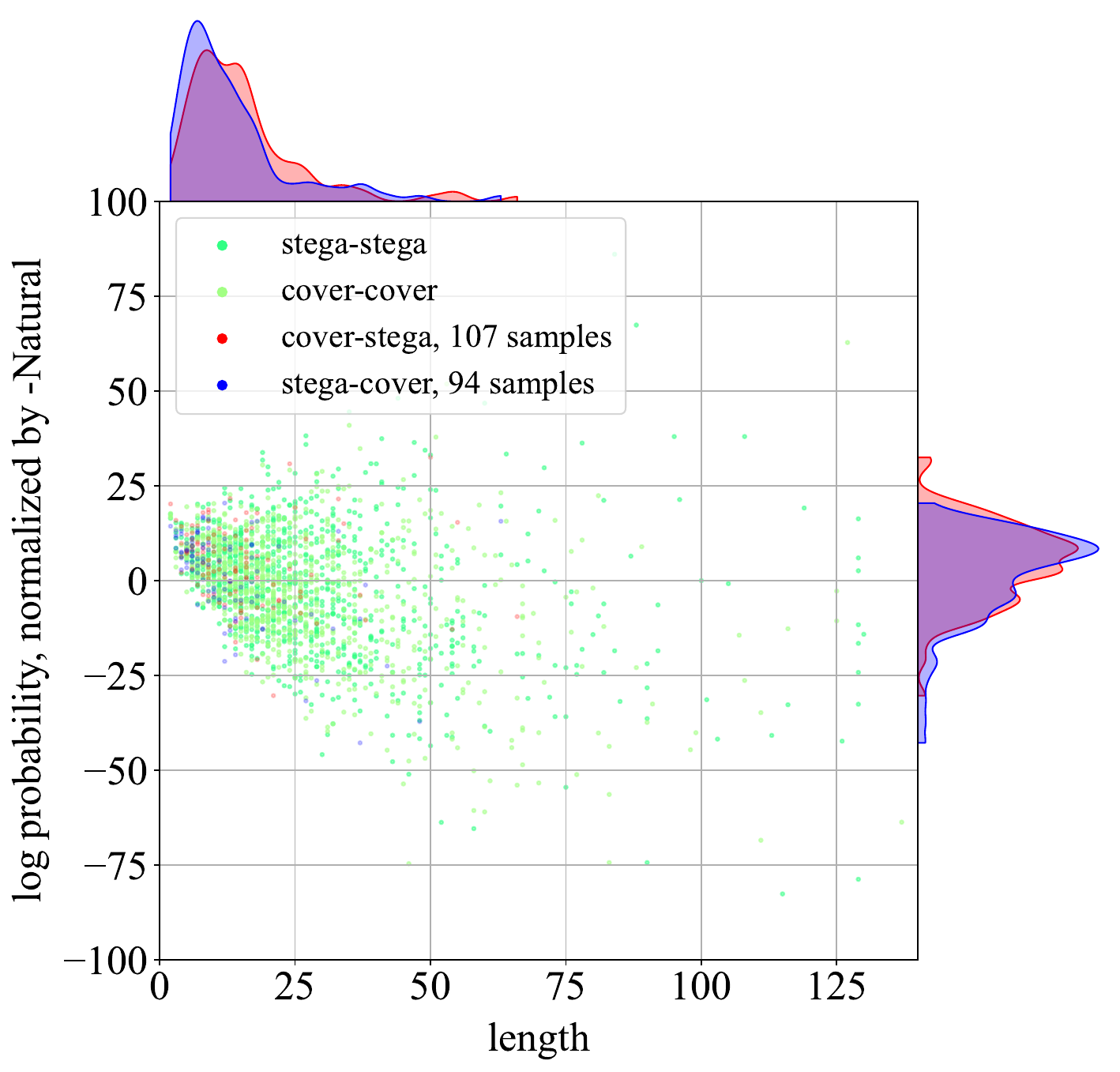} 
\label{Movie-ADG error distribution}}
\\
\subfigure[Distribution of error samples in News-AC]{
\includegraphics[width=0.31\textwidth]{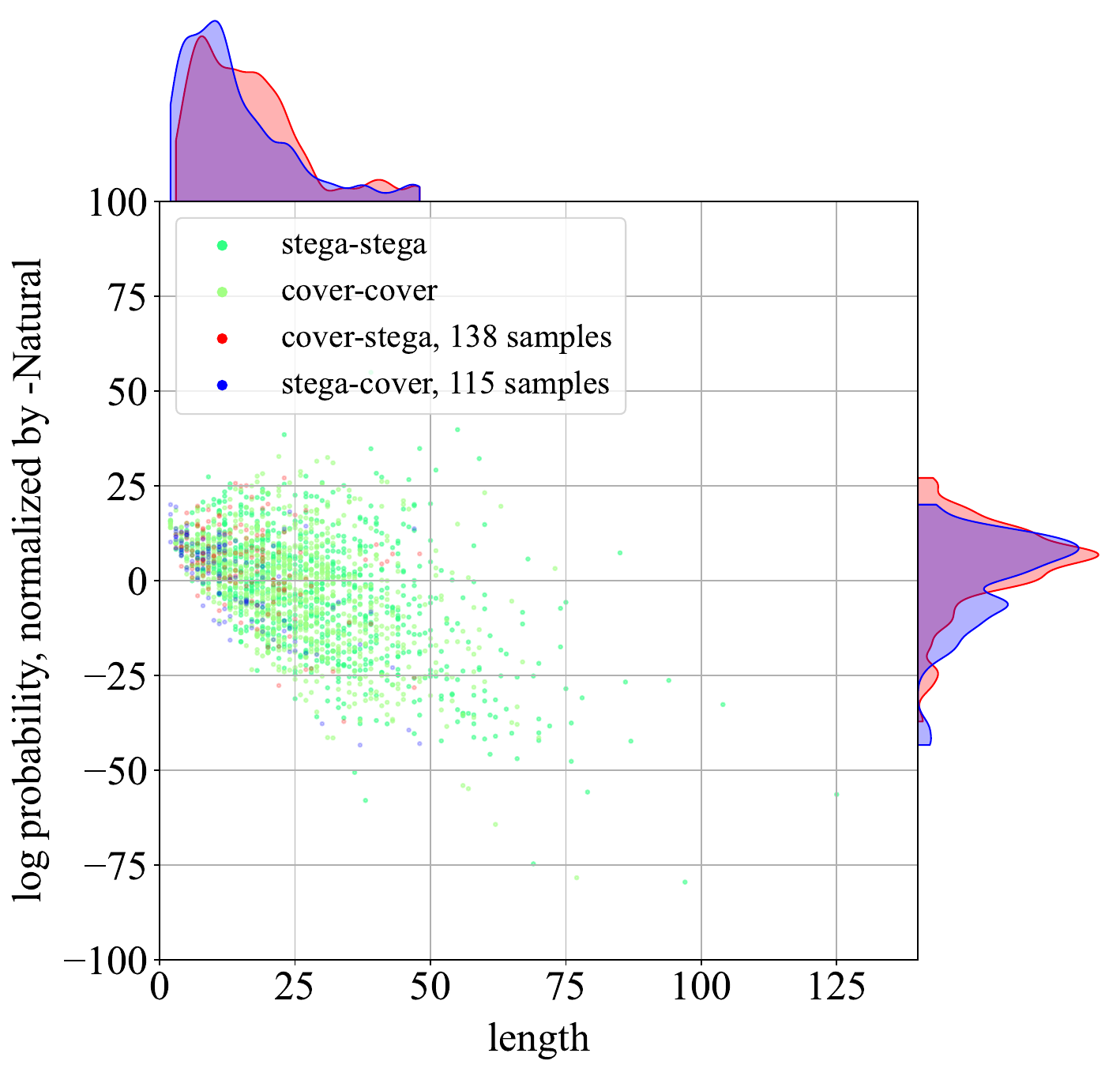}
\label{News-AC error distribution}}
\subfigure[Distribution of error samples in News-HC]{
\includegraphics[width=0.31\textwidth]{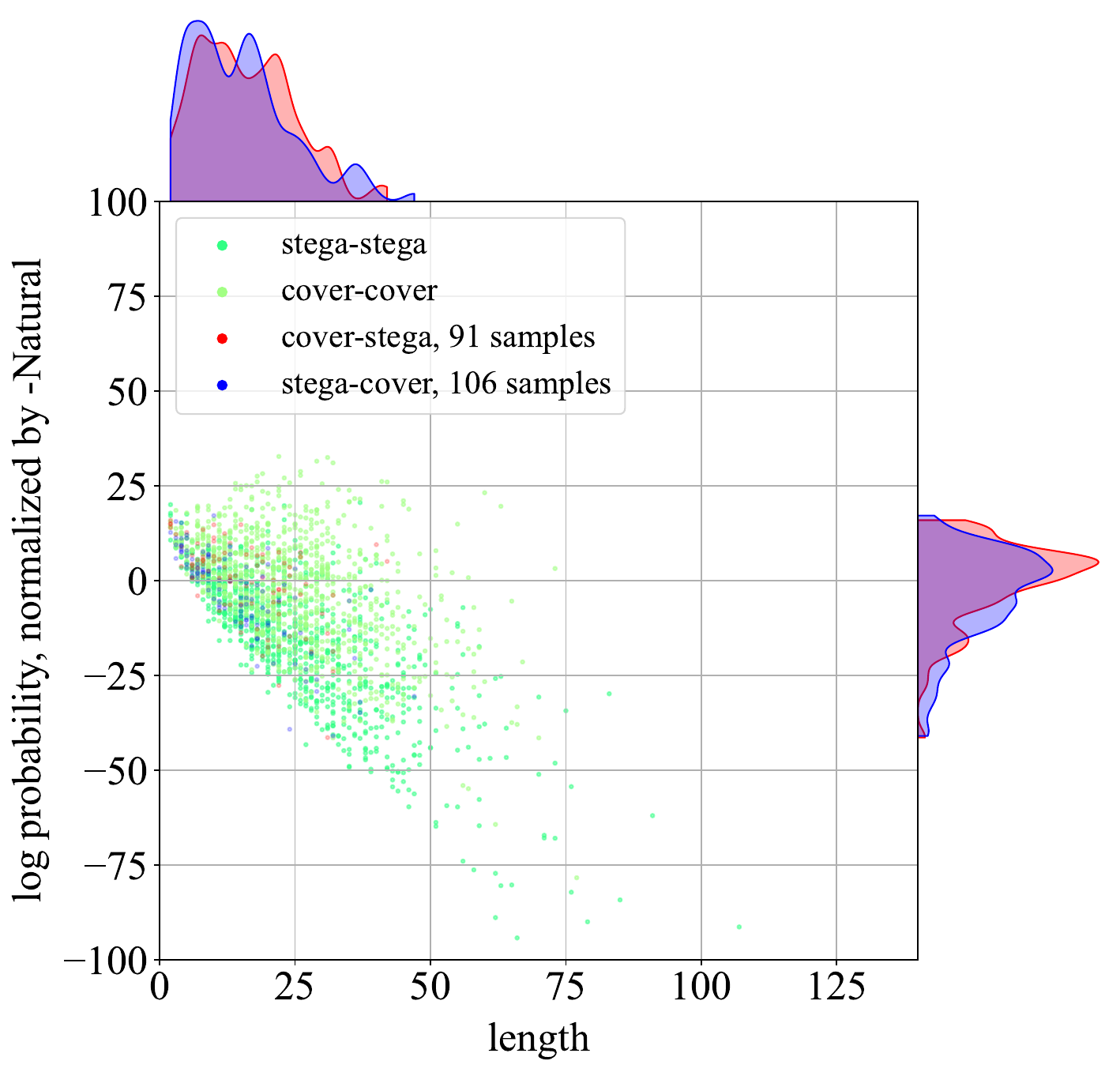}
\label{News-HC error distribution}}
\subfigure[Distribution of error samples in News-ADG]{
\includegraphics[width=0.31\textwidth]{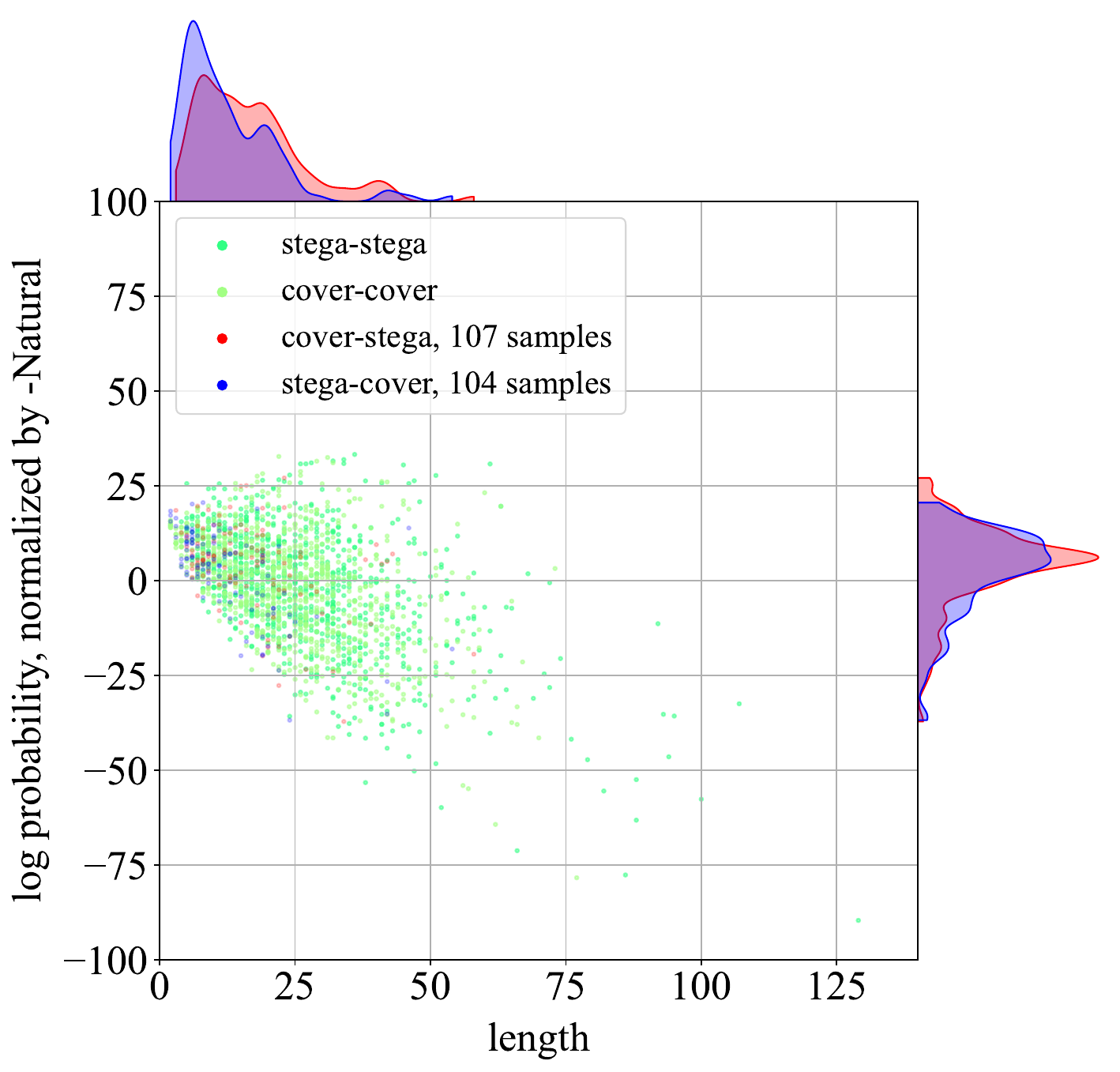} 
\label{News-ADG error distribution}}
\\
\subfigure[Distribution of error samples in Tweet-AC]{
\includegraphics[width=0.31\textwidth]{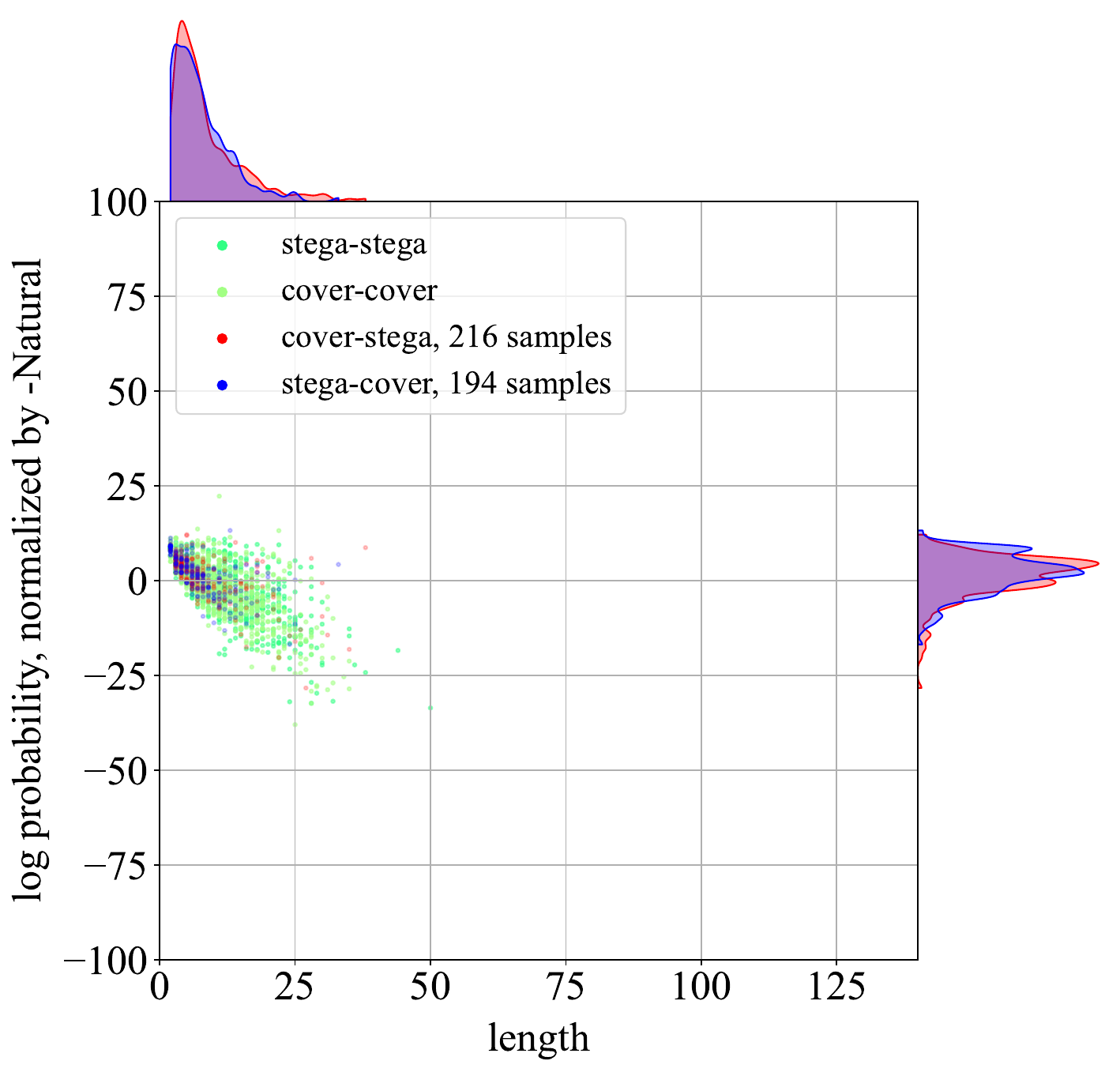}
\label{Tweet-AC error distribution}}
\subfigure[Distribution of error samples in Tweet-HC]{
\includegraphics[width=0.31\textwidth]{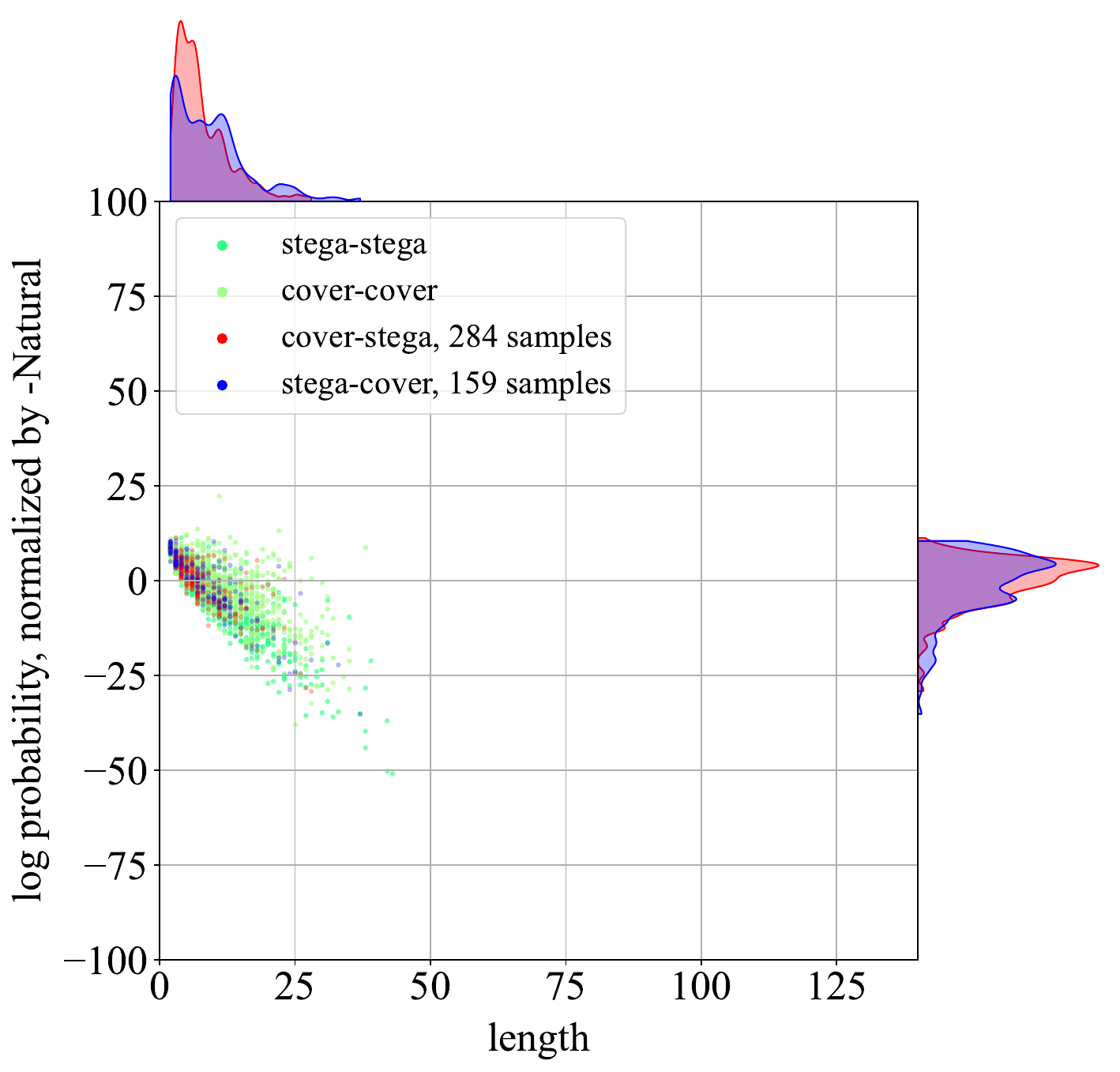}
\label{Tweet-HC error distribution}}
\subfigure[Distribution of error samples in Tweet-ADG]{
\includegraphics[width=0.31\textwidth]{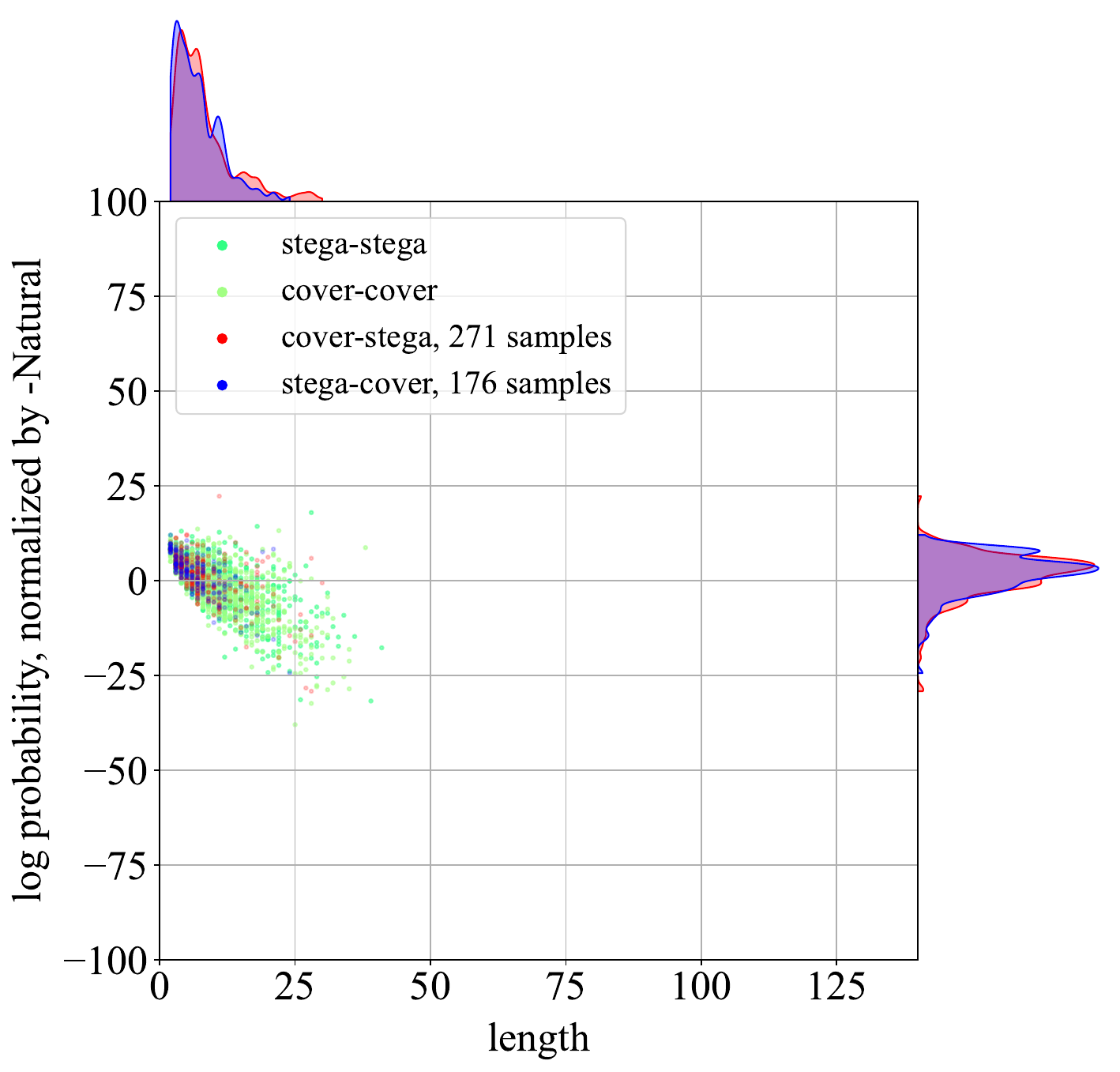} 
\label{Tweet-ADG error distribution}}
\caption{Distributions of error samples in domain-specific steganalysis. Each figure consists of the main scatter
plot of joint distribution and lines of marginal distributions of error samples at top and right. }
\label{Error distributions}
\end{figure*}

In Llama's experiments, it was noted that in particular datasets (Movie-AC, Movie-ADG, News-HC, News-ADG) training for 3 epochs led to superior outcomes than training for 10 epochs. This phenomenon could be attributed to the repetition of training data. Our models' results of training for 1-10 epochs on the Movie-AC dataset are illustrated in Fig. \ref{Different Epoches}. The figure indicates that Bloomz and Llama outperformed the BERT baseline models with just one epoch of training. Llama attained its highest accuracy in the fifth epoch, whereas it displayed a decreasing trend in accuracy with further training. On the other hand, Bloomz showed a gradual enhancement in accuracy over 10 epochs. Overall, Llama exhibited a detection accuracy that was about 3-5\% higher than Bloomz. 



\begin{figure*}[ht]
\subfigure[Detection accuracy of Bloomz on different length texts.]{\includegraphics[width=0.23\textwidth]{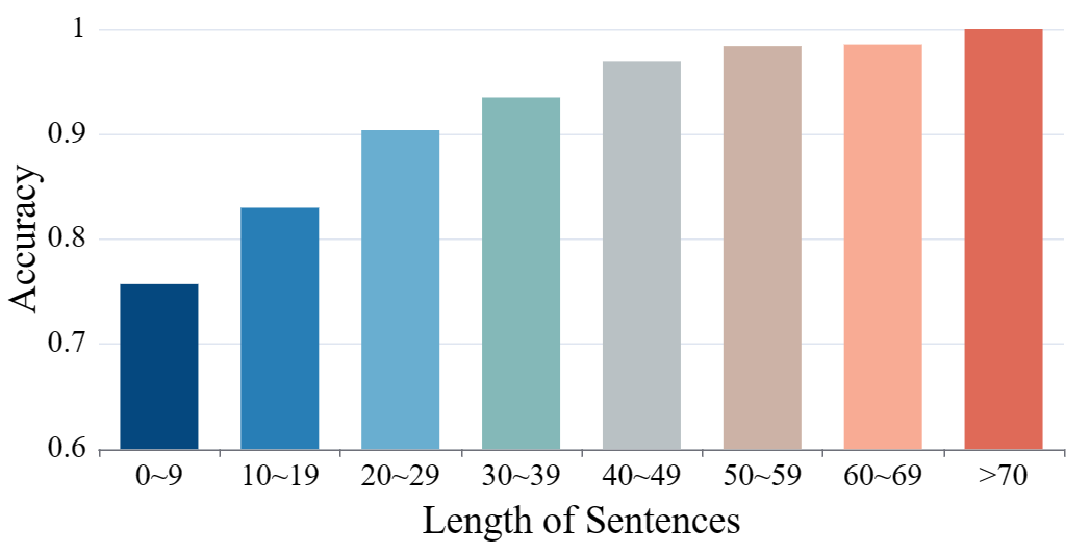}
\label{Different Length Bloomz}}
\subfigure[Detection accuracy of Llama on different length texts.]{
\includegraphics[width=0.23\textwidth]{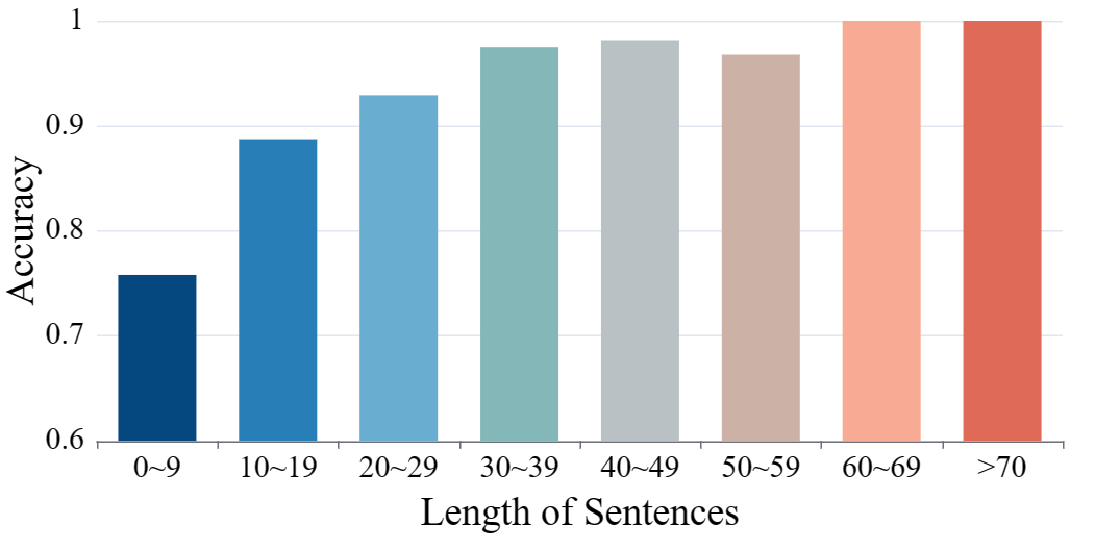}
\label{Different Length Llama}
}
\subfigure[Detection accuracy of Bloomz on texts of less than 10 tokens.]{
\includegraphics[width=0.23\textwidth]{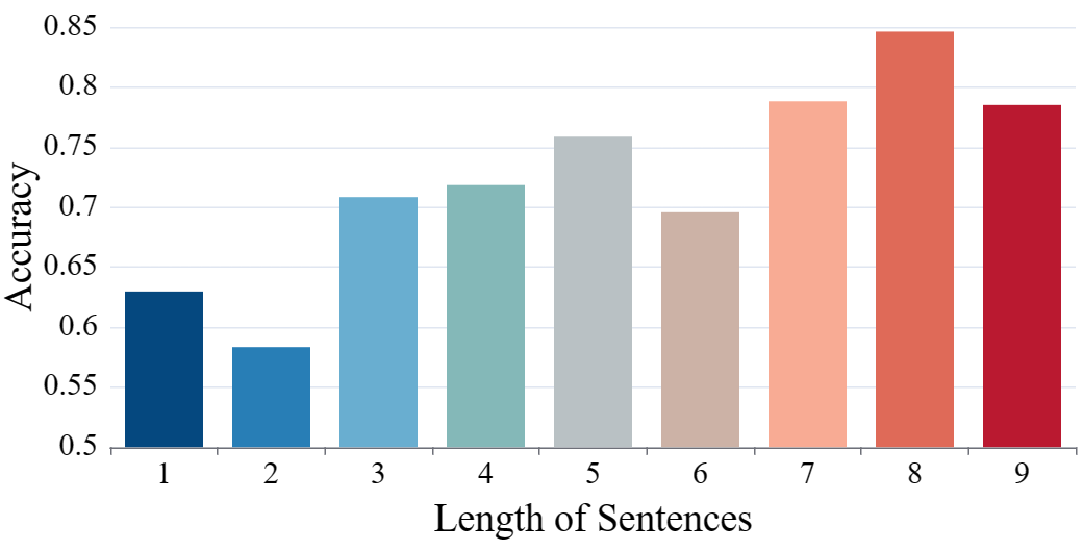}
\label{Shortest Sentences Bloomz}
}
\subfigure[Detection accuracy of Llama on texts of less than 10 tokens.]{
\includegraphics[width=0.23\textwidth]{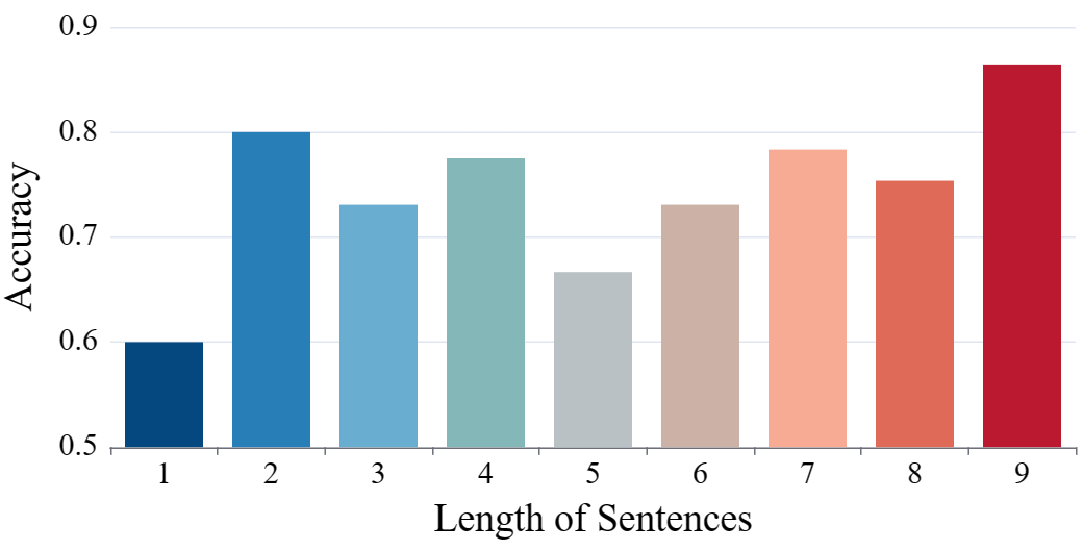}
\label{Shortest Sentences Llama}
}
\caption{Detection accuracy of Bloomz/Llama on sentences of various lengths.}
\end{figure*}
\begin{figure*}[ht]
\subfigure[Different Algorithms, Bloomz]{
\includegraphics[width=0.3\textwidth]{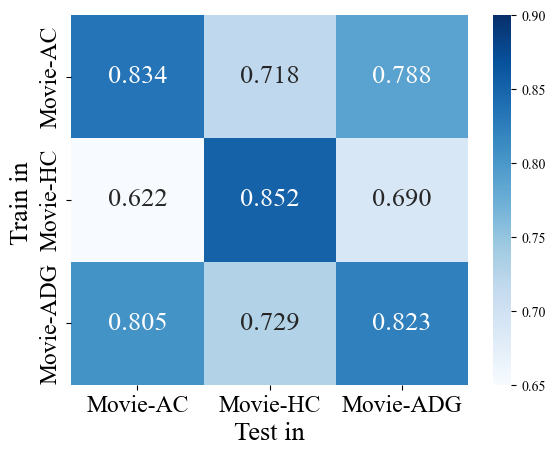}
\label{Transferability of bloomz, algorithms}}
\hspace{0.1cm}
\subfigure[Different Algorithms, Llama]{
\includegraphics[width=0.3\textwidth]{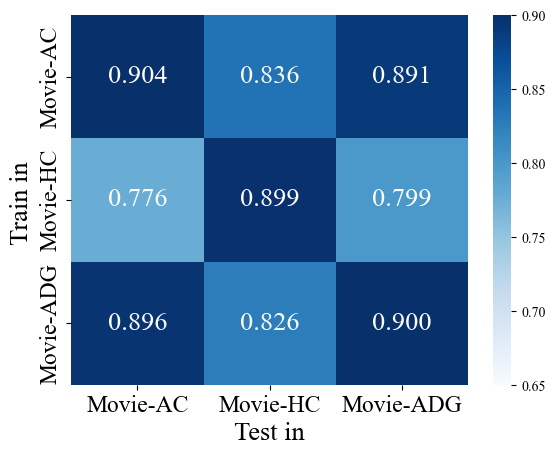}
\label{Transferability of llama, algorithms}}
\hspace{0.1cm}
\subfigure[Different Algorithms, best baseline]{
\includegraphics[width=0.3\textwidth]{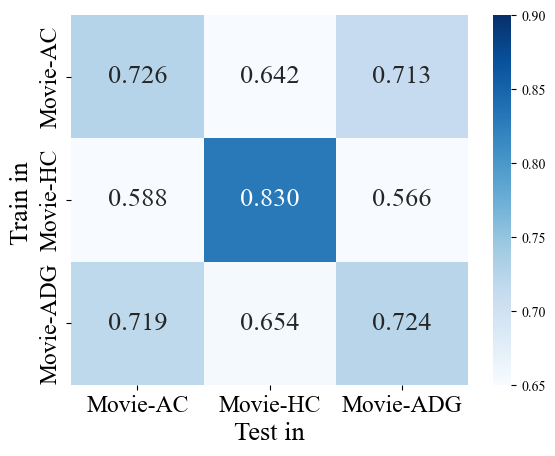} \label{Transferability of traditional models, algorithms}}
\\
\subfigure[Different Sources, Bloomz]{
\includegraphics[width=0.3\textwidth]{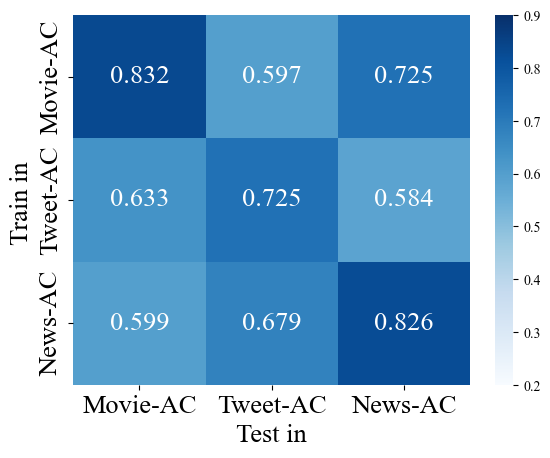}
\label{Transferability of bloomz, sources}}
\hspace{0.1cm}
\subfigure[Different Sources, Llama]{
\includegraphics[width=0.3\textwidth]{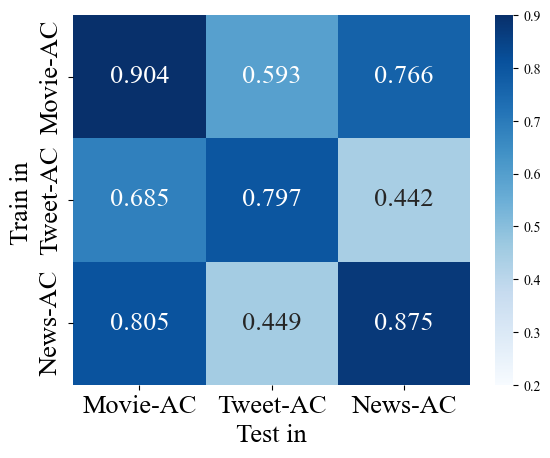} \label{Transferability of llama, sources}}
\hspace{0.1cm}
\subfigure[Different Sources, best baseline]{
\includegraphics[width=0.3\textwidth]{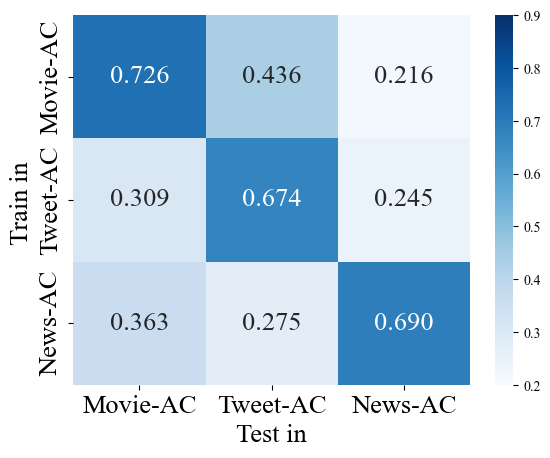} \label{Transferability of traditional models, sources}}
\caption{F1 scores of Bloomz, Llama and baseline mpdels on the domain-agnostic steganalysis task. y-axis denotes the training dataset and x-axis denotes the test dataset.} 
\label{Transferability of all models}
\end{figure*}

\subsection{Error Analysis of Domain-Specific Steganalysis}

To analyze the sentences that are non-detected (stegos that are not detected by our models) and incorrectly detected (covers that are wrongly detected as stegos by our models) in the domain-specific tasks, we scatter their distributions in Fig. \ref{Error distributions}.


In most cases, the normalized probability of the error samples is similar to the -Natural dataset, since their marginal distribution of normalized probability is concentrated near 0. The lengths of these samples are shorter than the average of -Natural dataset. Precisely, we find that for sentences of the same length, the lower the PPL the more likely it is to be misclassified by the model. From the distributions, it can be observed that a large proportion of texts is concentrated in a small region, and the error samples are located at the edge of this region. This phenomenon reminds us that these short and extremely fluent stegos are challenging samples for our models, which also proves that our models take fluency and rationality of text as standards of judgement. 
Most of the results show that the incorrectly detected samples are more than non-detected samples, which might be a warning sign of over-fitting. 

The detection accuracy of our models varies based on lengths of sentences. Fig. \ref{Different Length Bloomz} and \ref{Different Length Llama} show that sentences with less than 10 tokens have a detection rate of less than 80\%. However, sentences with over 20 tokens have an accuracy of 90\% or more. Additionally, sentences with more than 70 tokens have a detection rate of 100\%, suggesting that our models are nearly faultless in those cases.

\begin{table}[t]
\centering
\setlength{\tabcolsep}{3pt}
\caption{Short tweet samples. The rightmost column represents the detection results of a hybrid expert model consisting of baseline methods \cite{GNN,TS-CSW,TS-RNN,zou2020high,yang2019fast}. }
\setlength{\belowrulesep}{0pt}
\begin{tabular}{c|ccc}
\toprule
\midrule
Sentence & True label & Pred. of ours & Pred. of \cite{GNN,TS-CSW,TS-RNN,zou2020high,yang2019fast}\\
\hline
Yep thats a good thing & stego & stego & cover\\
Bit worried my poor pup & stego & cover & cover\\
Lol & cover & cover & cover\\
No I did not get it & cover & cover & stego\\
Yay & cover & stego & cover\\
\hline
\bottomrule

\end{tabular}
\label{Short tweet samples}
\end{table}
There are some short samples presented in Table \ref{Short tweet samples}. It is almost impossible for humans to differentiate between these fragmented but authentic stegos and covers, and the mix-up of baseline methods is also struggling. The broad range of tweets does help to generate seemingly innocent steganographic texts, and the restricted length of tweets presents a greater obstacle for detection models.


    


\begin{table*}[th]
\centering
\setlength{\belowrulesep}{0pt}
\caption{Result of different prompts, training for 3epoches in Tweet-AC. For the 7th prompt, our models generates non-standard outputs and does not perform any ability of steganalysis.}
\begin{tabular}{l|cccc|cccc}
\toprule
\midrule
\multirow{2}{*}{Prompts} & \multicolumn{4}{c|}{Bloomz-7B1} & \multicolumn{4}{c}{Llama-7B}\\
 & Accuracy & F1 score & Precision & Recall & Accuracy & F1 score & Precision & Recall\\
\hline
        Prompt 1  & 0.725 & 0.701 & \textbf{0.768} & 0.645 & 0.795 & 0.797  & 0.789 &	0.806 \\ 
        Prompt 2 & 0.741 & 0.748 & 0.728 & 0.769 & 0.803 & \textbf{0.808}  & 0.787 & \textbf{0.831} \\
        Prompt 3 & \textbf{0.751} & \textbf{0.761} & 0.732 & \textbf{0.792} & 0.803 & \textbf{0.808}  & 0.787 & \textbf{0.831} \\
        Prompt 4  & 0.688 & 0.658 & 0.727 & 0.600 & \textbf{0.805} & 0.807 & \textbf{0.798} & 0.816 \\ 
        Prompt 5  & 0.500 & 0.000 & 0.500 & 0.000 & 0.799 & 0.804 & 0.784 & 0.826 \\ 
        Prompt 6  & 0.704 & 0.697 & 0.714 & 0.680 & 0.800 & 0.801 & 0.796 & 0.805 \\ 
        Prompt 7  & - & - & - & - & - & - & - & - \\ 
        Prompt 8  & 0.733 & 0.744 & 0.715 & 0.775 & 0.789 & 0.788 & 0.791 & 0.786 \\ \hline
        \bottomrule
\end{tabular}
\label{prompts}
\end{table*}

\subsection{Domain-Agnostic Steganalysis}

We evaluated the domain-agnostic capability of two LLMs across various steganographic algorithms and datasets. This task consisted of training on one coding algorithm and testing on two other algorithms, or training on one type of dataset and testing on other datasets.
Each experiment entailed 3 epochs of training on the dataset. The outcomes are displayed in Fig. \ref{Transferability of bloomz, algorithms}, \ref{Transferability of bloomz, sources}, \ref{Transferability of llama, algorithms}, \ref{Transferability of llama, sources}. All of the baseline methods mentioned above were also tested, and Fig. \ref{Transferability of traditional models, algorithms}, \ref{Transferability of traditional models, sources} show the highest F1 scores achieved by these models. 


Our experiment has demonstrated the superiority of our models in terms of its domain-agnostic ability compared with baseline methods, surpassing the performance of baseline models considerably. As a result, we can assume that our models can robustly derive more accurate steganographic characteristics from various sources of text generated by different encoding algorithms. These extracted features can be applied to a wide range of steganographic texts. The distribution of text across various datasets presents a significant challenge to our models' domain-agnostic capability. This is due to certain data being classified as steganographic in the {Movie} dataset but not necessarily so in the {Tweet} dataset. Such variations must be taken into account to ensure optimal performance.

Furthermore, the pattern of Bloomz in the transfer experiments on different sources deviates significantly from that of the standard models. The {Movie} dataset has an average of 25.63 tokens per sentence and a perplexity value of 1059.65, whereas the {News} dataset has an average of 23.1 tokens per sentence and a perplexity value of 958. Despite having similar sentence lengths and perplexity values, the content of these two datasets is quite different. Bloomz performs significantly better when trained on the Movie-AC dataset and tested on the News-AC dataset, in contrast to the baseline models. Comparable results can be observed from the {Movie} and {Tweet} datasets, indicating that our model has learned unique steganographic features in comparison to the baseline models. According to Bloomz, the {News} dataset may be more comparable to the {Movie} dataset, whereas the {Tweet} dataset significantly differs from the {Movie} dataset. Conversely, the Llama views the {News} and {Movie} datasets as having minimal differences.  Llama exhibits robust transfer performance between the encoding algorithms, with negligible performance loss between the AC and ADG algorithms. Nonetheless, it encounters considerable performance degradation in transfer tasks across datasets, potentially attributable to the restricted diversity of training data implemented for the Llama model.

However, the {Tweet} dataset exhibits notably weaker transfer performance than the majority of datasets. As we mentioned before, part of our models' ability is from the perception of text fluency and rationality. While the sentences in Tweet dataset are often fragmented spoken expressions. We assume that the reason can be attributed to the complexity and non-normative nature of Twitter texts.


\begin{table}[t]
  \centering
  \caption{Prompts}
  \setlength{\aboverulesep}{2pt}
  \setlength{\belowrulesep}{2pt}
    \begin{tabular}{c|llllllllll}
    \toprule
    \hline
    \multirow{5}[0]{*}{Prompt 1} & \multicolumn{10}{l}{\multirow{5}[0]{*}{\makecell[l]{\#\#\# Text: \{input sentence\} \\ \#\#\# Question: Is the above text steganographic \\ or non-steganographic?  \\ \#\#\# Answer: \{input label (steganographic or \\ non-steganographic)\} }}} \\
          & \multicolumn{10}{l}{} \\
          & \multicolumn{10}{l}{} \\
          \\
          
          \\\hline
    \multirow{3}[0]{*}{Prompt 2} & \multicolumn{10}{l}{\multirow{3}[0]{*}{\makecell[l]{\#\#\# Text: \{input sentence\}\\
    \#\#\# Question: Is the above text steganographic?\\
    \#\#\# Answer: \{input label (yes or no)\}}}} \\
          & \multicolumn{10}{l}{} \\
          & \multicolumn{10}{l}{} \\\hline
    \multirow{3}[0]{*}{Prompt 3} & \multicolumn{10}{l}{\multirow{3}[0]{*}{\makecell[l]{\#\#\# Here is a tweet: \{input sentence\}\\
    \#\#\# Question: Is the above tweet steganographic?\\
    \#\#\# Answer: \{input label (yes or no)\}}}} \\
          & \multicolumn{10}{l}{} \\
          & \multicolumn{10}{l}{} \\\hline
    \multirow{3}[0]{*}{Prompt 4} & \multicolumn{10}{l}{\multirow{3}[0]{*}{\makecell[l]{\#\#\# \{input sentence\}\\\#\#\# \{input label (steganographic or\\ non-steganographic)\}}}} \\
          & \multicolumn{10}{l}{} \\
          & \multicolumn{10}{l}{} \\\hline
    \multirow{3}[0]{*}{Prompt 5} & \multicolumn{10}{l}{\multirow{3}[0]{*}{\makecell[l]{\#\#\# \{input sentence\}\\\#\#\# \{input label (0 or 1)\}}}} \\
          & \multicolumn{10}{l}{} \\
          & \multicolumn{10}{l}{} \\\hline
    \multirow{3}[0]{*}{Prompt 6} & \multicolumn{10}{l}{\multirow{3}[0]{*}{\makecell[l]{\#\#\# Text: \{input sentence\}\\\#\#\# Label: \{input label (steganographic or \\ non-steganographic)\}}}} \\
          & \multicolumn{10}{l}{} \\
          & \multicolumn{10}{l}{} \\\hline
    \multirow{3}[0]{*}{Prompt 7} & \multicolumn{10}{l}{\multirow{3}[0]{*}{\makecell[l]{\#\#\# Text: \{input sentence\}\\\#\#\# Label: \{input label (stega or cover)\}}}} \\
          & \multicolumn{10}{l}{} \\
          & \multicolumn{10}{l}{} \\\hline
    \multirow{3}[0]{*}{Prompt 8} & \multicolumn{10}{l}{\multirow{3}[0]{*}{\makecell[l]{\#\#\# Text: \{input sentence\}\\\#\#\# Label: \{input label (0 or 1)\}}}} \\
          & \multicolumn{10}{l}{} \\
          & \multicolumn{10}{l}{} \\\hline
    \bottomrule
    \end{tabular}%
  \label{Prompts}%
\end{table}%
\begin{table*}[ht]
  \centering
  \setlength{\tabcolsep}{10pt}
  \renewcommand{\arraystretch}{1.0}
  \caption{generic steganalysis result. The rightmost two columns represent the result of the best baseline method trained in these datasets, while our model is only tested. }
    \begin{tabular}{cc|cccccc|cc}
    \toprule
    \hline
    \multicolumn{2}{c|}{Datasets \&} & \multicolumn{6}{c|}{GS-Llama} & \multicolumn{2}{c}{Best Baseline} \\
    \multicolumn{2}{c|}{Algorithms} & TP & FN & FP & TN &
    Accuracy & F1 score & Accuracy & F1 score \\
    \hline
    \multirow{4}[0]{*}{Movie} & \multicolumn{1}{c|}{AC} & 935   & 65    & 53    & 947   & 94.10 & 94.06 & \multicolumn{1}{c}{68.17} & \multicolumn{1}{c}{67.92} \\
          & \multicolumn{1}{c|}{HC} & 839   & 161   & 53    & 947   & 89.30 & 88.69 & \multicolumn{1}{c}{79.05} & \multicolumn{1}{c}{80.23} \\
          & \multicolumn{1}{c|}{ADG} & 898   & 102   & 53    & 947   & 92.25 & 92.06 & \multicolumn{1}{c}{71.80} & \multicolumn{1}{c}{68.81} \\
          
          & DISCOP & 761   & 238   & 53    & 947   & \textbf{85.40} & \textbf{83.90} & 76.15     & 75.84\\
          \hline
    \multirow{3}[0]{*}{News} & \multicolumn{1}{c|}{AC} & 744   & 256   & 124   & 876   & \textbf{81.00} & \textbf{79.66} & \multicolumn{1}{c}{70.25} & \multicolumn{1}{c}{68.28} \\
          & \multicolumn{1}{c|}{HC} & 669   & 331   & 124   & 876   & 77.25 & 74.62 & \multicolumn{1}{c}{77.90} & \multicolumn{1}{c}{79.64} \\
          & \multicolumn{1}{c|}{ADG} & 757   & 243   & 124   & 876   & \textbf{81.65} & \textbf{80.49} & \multicolumn{1}{c}{70.00} & \multicolumn{1}{c}{67.62} \\
          \hline
    \multirow{3}[0]{*}{Tweet} & \multicolumn{1}{c|}{AC} & 770   & 230   & 217   & 783   & 77.65 & 77.50 & \multicolumn{1}{c}{66.02} & \multicolumn{1}{c}{64.24} \\
          & \multicolumn{1}{c|}{HC} & 818   & 182   & 217   & 783   & 80.05 & 80.39 & \multicolumn{1}{c}{71.30} & \multicolumn{1}{c}{73.28} \\
          & \multicolumn{1}{c|}{ADG} & 787   & 213   & 217   & 783   & 78.50 & 78.54 & \multicolumn{1}{c}{71.80} & \multicolumn{1}{c}{68.81} \\
          \hline
    \multirow{3}[0]{*}{Commonsense} & \multicolumn{1}{c|}{AC} & 710   & 290   & 225   & 775   & \textbf{74.25} & \textbf{73.39} & \multicolumn{1}{c}{60.30} & \multicolumn{1}{c}{59.57} \\
          & \multicolumn{1}{c|}{HC} & 762   & 238   & 225   & 775   & \textbf{76.85} & \textbf{76.70} & \multicolumn{1}{c}{67.40} & \multicolumn{1}{c}{65.50} \\
          & \multicolumn{1}{c|}{ADG} & 688   & 312   & 225   & 775   & \textbf{73.15} & \textbf{71.93} & \multicolumn{1}{c}{59.65} & \multicolumn{1}{c}{57.26} \\
          \hline
    \multirow{3}[0]{*}{Aclimdb} & \multicolumn{1}{c|}{AC} & 920   & 80    & 33    & 967   & \textbf{94.35} & \textbf{94.21} & \multicolumn{1}{c}{85.50} & \multicolumn{1}{c}{85.50} \\
          & \multicolumn{1}{c|}{HC} & 978   & 22    & 33    & 967   & \textbf{97.25} & \textbf{97.30} & \multicolumn{1}{c}{92.25} & \multicolumn{1}{c}{92.23} \\
          & \multicolumn{1}{c|}{ADG} & 986   & 14    & 33    & 967   & \textbf{97.65} & \textbf{97.78} & \multicolumn{1}{c}{78.75} & \multicolumn{1}{c}{78.14} \\
          \hline
    \bottomrule
    \end{tabular}%
  \label{tab: generic steganalysis result}%
\end{table*}%

\subsection{Effectiveness of Prompts}
\label{Effectiveness of different prompts}

To assess the impact of prompts on models' detection performance, including inputting instruction options (separators, hints and questions) and outputting format options (such as long terms like ``steganographic/non-steganographic", short terms like ``stega/cover", simple responses like ``yes/no", or even numerical answers like ``0/1"), we employed the prompts presented in Table \ref{Prompts}.


As an example, we tested the effect of these prompts in {Tweet+AC} dataset, and the results are shown in Table \ref{prompts}.

In the prompts containing a question (Prompt 1,2,3), the detection performance for the ``Yes/No" labels is better than the performance for the ``Steganographic/Non-steganographic" labels. The efficiency of Bloomz is further improved by incorporating the textual category information (``tweet") in the input prompt. However, there are some minor performance changes for Llama.

For the prompts with only separators (Prompts 4 and 5), the effectiveness of Prompt 4 for generating long words is similar to that of the full prompt. On the other hand, Prompt 5, which outputs either 0 or 1, completely destroys the steganalysis capability of Bloomz, while Llama is still able to maintain a robust result.

Regarding the results from prompts 6, 7, and 8, we can verify that the most suitable output for LLMs is to be linguistically precise, whereas meaning of ``cover" is excessively broad, resulting in model outputs that deviate from our expectations. The difference in detection performance between numeric and long words is due to the model's understanding of steganographic/non-steganographic concepts. Owing to the resemblance of ``Steganographic" and ``Non-steganographic", the model produces lower losses for this output. Nevertheless, the use of these low-loss labels does not result in optimal detection performance.

To summarise, we anticipate that all inputs and outputs closely emulate natural language.
Additionally, these inputs should be rational and unambiguous for LLMs to comprehend easily. Ideally, the output labels should display considerable variation at the token level, have a unique semantic interpretation and conflicting meanings.

\subsection{Towards General Steganalysis}

After the sufficient experiments presented above, we found that there are 2 main challenges for LLMs in the steganalysis task. The first challenge is that short texts are more difficult to detect, which is also quite problematic for humans and baseline methods. The second challenge is that as the degree of text diversity increases, the decline in detection accuracy is difficult to compensate for. 


With sufficient experimental preparation, now it's time to train a generic steganalysis LLM. First, in terms of model training settings, we choose Llama-7B training for 3 epochs with Prompt 2. Then for the training dataset, we extract 10000 sentences from Movie-AC and 5000 sentences from Tweet-HC as steganographic dataset, along with 10000 sentences from Movie-Natural and 5000 sentences from Tweet-Natural as natural dataset. From the results of domain-agnostic task it can be observed that the Movie dataset contains various texts and the distribution of this dataset is the most spread out, which enables our model to learn a diverse range of textual features. Thus we choose the Movie dataset as the basis. To enhance the models' ability to detect short text, we chose a part of the Tweet dataset as supplementary. Under these settings, our model took 9 hours to train on one RTX3090 GPU.

Besides the Movie and Tweet datasets, we test our model in new datasets (Aclimdb and Commonsense) and a new steganographic algorithm (DISCOP). Since we do not use any data from the News dataset, the results on the News dataset also demonstrate the capability of general detection of our model. The main results of general steganalysis task are shown in Table \ref{tab: generic steganalysis result}.

In most cases, the detection rate and F1-score of our model are much higher than the baseline method that was trained and tested in the same dataset, with an increase in detection accuracy by approximately 10\%. Compared with the results of domain-agnostic steganalysis that only uses the Movie-AC dataset, our model has significant gains. 
In this experiment, some different patterns are exposed. In the datasets that do not participate in training, the incorrectly detected samples are explicitly less than non-detected samples, which confirms our conjectures about over-fitting. We observed that in these datasets the incorrectly detected samples are shorter and with higher PPL, compared with the non-detected samples. Therefore, for the generality of the model, constructing a diverse dataset is significant for model training.

We will continue to improve this general steganalysis model, aiming at training a practical LLM for steganalysis.

\section{Conclusion}


In this paper, we proposed a novel steganalysis method that different from the current mainstream classification-based methods. To the best of our knowledge, we are the first to construct a generation-based steganalysis method with LLMs. To directly generate the readable detection output without any additional deep learning module, we fine-tuned LLMs with LoRA framework, maintaining the original capability of LLMs. Our method showed superior performance in different scenarios compared to all existing steganalysis methods.
Results show that the detection ability of our method is based on the fluency and rationality of the text, only a minor proportion of short stegos with low PPL are likely to escape the detection. Furthermore, our models demonstrated exceptional capability in domain-agnostic tasks, providing robust detection results even when trained and tested on different datasets. 
This infers that LLMs gained wider-ranging characteristics of steganographic text during instruction fine-tuning, which remain effective despite significant variations in the text.

With the above results, we finally fine-tuned a general steganalysis model that can be applied to various tasks. In most cases, our method gains higher detection accuracy when only tested in the dataset than baseline methods trained in the same dataset. 

We hope that our method will provide new insights for steganalysis in the era of LLMs. Our goal is to further improve the detection capabilities of our general steganalysis model and make it applicable to real-world situations.

\bibliographystyle{IEEEtran}
\bibliography{sample,IEEEabrv}

\end{document}